\documentclass[12pt]{article}
\usepackage{graphicx}
\usepackage{etoolbox}
\usepackage{hyperref}
\hypersetup{
    colorlinks=false,
    linkcolor=blue,
    filecolor=magenta,      
    urlcolor=cyan,
    }
\usepackage{times}
\usepackage{bm}
\usepackage{makecell}
\usepackage{color}
\usepackage{amssymb}
\usepackage{amsmath}
\usepackage{braket}
\usepackage[table]{xcolor}
\usepackage{soul}
\usepackage{multirow}
\usepackage{float}

\usepackage[resetlabels]{multibib}
\newcites{Methods}{Methods only references}

\newenvironment{sciabstract}{%
\begin{quote} \bf}
{\end{quote}}

\usepackage{graphicx}

\graphicspath{ {figures/} }

\makeatletter
\let\saved@includegraphics\includegraphics
\AtBeginDocument{\let\includegraphics\saved@includegraphics}
\renewenvironment*{figure}{\@float{figure}}{\end@float}
\makeatother

\makeatletter
\newcommand*{\centerfloat}{%
  \parindent \z@
  \leftskip \z@ \@plus 1fil \@minus \textwidth
  \rightskip\leftskip
  \parfillskip \z@skip}
\makeatother

\usepackage{lineno} 
\begin{document}
% \linenumbers
\setstcolor{red}

\title{Probing many-body dynamics in a two dimensional dipolar spin ensemble}

\author{
E.J.~Davis,$^{1,*}$
B.~Ye,$^{1,*}$
F.~Machado,$^{1,2,*}$

S.A.~Meynell,$^{3}$

W.~Wu,$^{1,2}$

T.~Mittiga,$^{1,2}$\\

W.~Schenken,$^{3}$
M.~Joos,$^{3}$

B.~Kobrin,$^{1,2}$
Y.~Lyu,$^{1}$

Z.~Wang,$^{1,2}$
D.~Bluvstein,$^{4}$\\

S.~Choi,$^{1}$
C.~Zu,$^{1,2,5}$
A.C.~Bleszynski Jayich,$^{3, \dag}$
N. Y. Yao$^{1,2,\ddag}$\\
\\
\normalsize{$^{1}$Department of Physics, University of California, Berkeley, CA 94720, USA}\\
\normalsize{$^{2}$Materials Science Division, Lawrence Berkeley National Laboratory, Berkeley, CA 94720, USA}\\
\normalsize{$^{3}$Department of Physics, University of California, Santa Barbara, CA 93106, USA}\\
\normalsize{$^{4}$Department of Physics, Harvard University, Cambridge, MA 02138, USA}\\
\normalsize{$^{5}$Department of Physics, Washington University, St. Louis, MO 63130, USA}\\

\normalsize{$^*$These authors contributed equally to this work.}\\
\normalsize{$^\dag$To whom correspondence should be addressed; E-mail:  ania@physics.ucsb.edu}\\
\normalsize{$^\ddag$To whom correspondence should be addressed; E-mail:  norman.yao@berkeley.edu}\\

}

\date{}

\baselineskip24pt
\maketitle

\vspace{5mm}

% \section*{Abstract}
\begin{sciabstract}
The most direct approach for characterizing the quantum dynamics of a strongly-interacting  system is to measure the time-evolution of its full many-body  state. Despite the conceptual simplicity of this approach, it quickly becomes intractable as the system size grows. An alternate framework is to think of the many-body dynamics as generating noise, which can be measured by the decoherence of a probe qubit. Our work centers on the following question: What can the decoherence dynamics of such a probe tell us about the many-body system? In particular, we utilize optically addressable probe spins to experimentally characterize both static and dynamical properties of strongly-interacting magnetic dipoles. Our experimental platform consists of two types of spin defects in diamond: nitrogen-vacancy (NV) color centers (probe spins) and substitutional nitrogen impurities (many-body system). We demonstrate that signatures of the many-body system's dimensionality, dynamics, and disorder are naturally encoded in the functional form of the NV's  decoherence profile. Leveraging these insights, we directly characterize the two-dimensional nature of a nitrogen delta-doped diamond sample. In addition, we explore two distinct facets of the many-body dynamics: First, we address a persistent debate about the microscopic nature of spin dynamics in strongly-interacting dipolar systems. Second, we demonstrate direct control over the correlation time of the many-body system. Finally, we demonstrate polarization exchange between NV and P1 centers, opening the door to quantum sensing and simulation using two-dimensional spin-polarized ensembles.
\end{sciabstract}

\maketitle
% \section*{Main text}
% \subsection*{Introduction}
Understanding and controlling the interactions between a single quantum degree of freedom and its environment represents a fundamental challenge within the quantum sciences~\cite{purcell1995spontaneous,lorenza1999dynamical, houck2008controlling, de2010universal, tyryshkin2012electron, klauder1962spectral, schweiger2001principles, kofman2000acceleration, romach2015spectroscopy}. Typically, one views this challenge through the lens of mitigating decoherence---enabling one to engineer a highly coherent qubit by decoupling it from the environment~\cite{lorenza1999dynamical, houck2008controlling, de2010universal, tyryshkin2012electron, kleppner1981inhibited, kotler2011single, bar2012suppression}. However, the environment itself may consist of a strongly-interacting many-body system, which naturally leads to an alternate perspective; namely, using the decoherence dynamics of the qubit to probe the fundamental properties of the many-body system~\cite{klauder1962spectral, schweiger2001principles, herzog1956transient, kubo2012statistical, salikhov1981theory, chiba1972electron,altman2004probing, hofferberth2008probing}.  Discerning the extent to which such ``many-body noise'' can provide insight into transport dynamics, low-temperature order, and generic correlation functions  of an interacting system remains an essential open question.%~\cite{chatterjee2019diagnosing,khoo2021universal}. 

The complementary goals of probing and eliminating many-body noise have motivated progress in magnetic resonance spectroscopy for decades, including seminal work exploring the decoherence of paramagnetic defects in solids~\cite{herzog1956transient, klauder1962spectral, kubo2012statistical, salikhov1981theory, chiba1972electron, schweiger2001principles, fel1996configurational}. 
More recently, many of the developed techniques have re-emerged in the study of solid-state spin ensembles containing optically-polarizable color centers. 
The ability to prepare spin-polarized pure states enables fundamentally new prospects in quantum science, from the exploration of novel phases of matter \cite{choi2017observation} to the development of new sensing protocols \cite{sushkov2014magnetic}.

Prospects for optically-polarizable spin ensembles in quantum sensing and simulation could be further enhanced by moving to two-dimensional systems, which represents a long-standing engineering challenge for the color-center community~\cite{ohno2012engineering, mclellan2016patterned, eichhorn2019optimizing}. Despite continued advances in fabrication, the stochastic nature of defect generation strongly constrains the systems one can create. The potential rewards are substantial enough to merit repeated engineering efforts: Two-dimensional, long-range interacting spin systems are known to host interesting ground state phases such as spin liquids~\cite{yao2018quantum,chomaz2019long,semeghini2021probing}. 
Moreover, two-dimensional spin ensembles enable improved sensing capabilities owing to increased coherence times and uniform distance from the target~\cite{SM}.

In this paper, we investigate many-body noise generated by a thin layer of paramagnetic defects in diamond. Specifically, we combine nitrogen delta-doping during growth with local electron irradiation to fabricate a diamond sample (S1) where paramagnetic defects are confined to a layer whose width is, in principle, much smaller than the average spin-defect spacing~[Fig.~\ref{fig:fig1}(a, b)]~\cite{ohno2012engineering, mclellan2016patterned, eichhorn2019optimizing}. This layer contains a hybrid spin system consisting of two types of defects: spin-1 nitrogen-vacancy (NV) centers and spin-1/2 substitutional nitrogen (P1) centers. The dilute NV centers can be optically initialized and read-out, making them a natural probe of the many-body noise generated by the strongly-interacting P1 centers. In addition, we demonstrate a complementary role for the NV centers, 
as a source of spin polarization for the optically-dark P1 centers; in particular, by using a Hartmann-Hahn protocol, we directly transfer polarization between the two spin ensembles.

We experimentally characterize the P1's many-body noise via the decoherence dynamics of NV probe spins. To elucidate our results, we first present a theoretical framework that unifies and generalizes existing work, predicting a non-trivial temporal profile that exhibits a crossover between two distinct stretched exponential decays (for the average coherence of the probe spins)  [Fig.~\ref{fig:fig1}]~\cite{anderson1953exchange, herzog1956transient, klauder1962spectral, fel1996configurational, kubo2012statistical, salikhov1981theory, chiba1972electron}. 
Beyond solid-state spin systems, the framework naturally extends to a broader class of quantum simulation platforms, including trapped ions, Rydberg  atoms, and ultracold polar molecules~\cite{georgescu2014quantum}.
Crucially, we demonstrate that the associated stretch powers contain a wealth of information about both the static and dynamical properties of the many-body spin system.

We focus on three such properties. First, the stretch power contains a direct signature revealing the dimensionality of the disordered many-body system. Unlike previous work on lower-dimensional ordered systems in magnetic resonance spectroscopy~\cite{engelsberg1973nuclear, cho1996h, cho2005multispin}, we cannot leverage conventional methods such as X-ray diffraction to characterize our disordered spin ensemble. To the best of our knowledge, studying the decoherence dynamics provides the only robust method to determine the effective dimensionality seen by the spins.

The stretch power of the NV centers' decoherence can also distinguish between different forms of spectral diffusion, shedding light
on the nature of local spin fluctuations.  In particular, we demonstrate that the P1 spin-flip dynamics are inconsistent with the conventional
expectation of telegraph noise, but rather follow that of
a Gauss-Markov process [Table~\ref{tab:table1}]. Understanding the statistical properties of the many-body noise and the precise physical settings
where such noise emerges remains the subject of active debate~\cite{mims1968phase, klauder1962spectral, chiba1972electron, abe2004electron,zhong2015optically, de2003theory, wang2013spin,bauch2020decoherence, hanson2008coherent,de2010universal}. %remains the subject of active debate~\cite{de_Sousa_2009}. 

Finally, the crossover in time between different stretch powers allows one to extract the many-body system's correlation time.
We demonstrate this behavior by actively controlling the correlation time of the P1 system via polychromatic driving, building upon techniques previously utilized in broadband decoupling schemes~\cite{Ernst1966}.

%%%%%%%%%%%%%%%%%%%%%%%%%%%%%%%%%%%%%%%%%%
%% THEORY SUBSECTION %%
%%%%%%%%%%%%%%%%%%%%%%%%%%%%%%%%%%%%%%%%%%
\subsection*{Theoretical framework for decoherence dynamics induced by many-body noise}
\begin{figure}[H]
    \centering
    \includegraphics[width=6.5in]{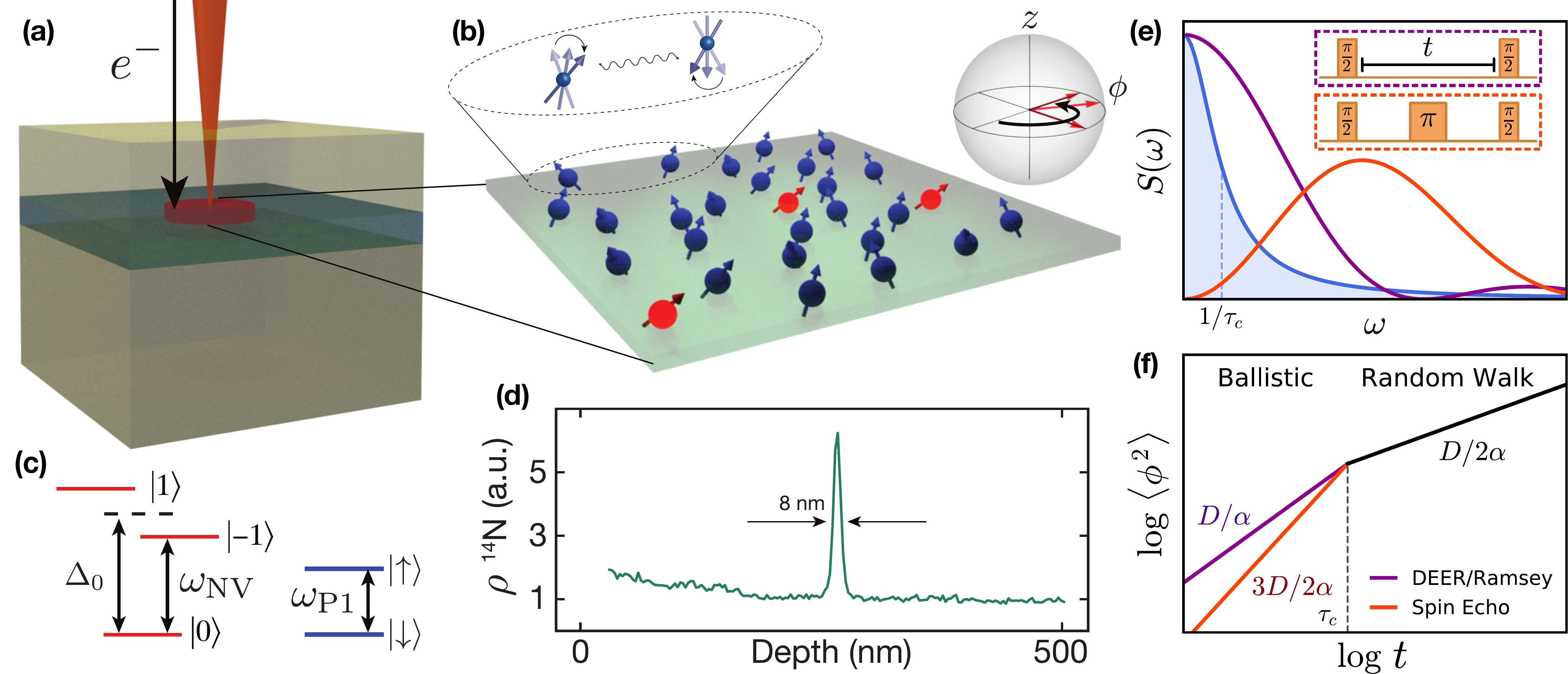}
    \caption{
    \textbf{Experimental platform and theoretical framework} (a) A delta-doped layer of $^{14}$N (green) is grown on a diamond substrate. NV centers are created via local electron irradiation (orange beam) and subsequent high-temperature annealing. 
    (b) Schematic depiction of a two-dimensional layer of NV (red) and P1 (blue) centers. Dilute NV centers function as probe spins of the dense, disordered P1 system.
    The P1s exhibit spin-flip dynamics driven by magnetic dipole-dipole interactions (zoom). 
    Ising interactions with the P1 system cause the NV to accumulate phase, $\phi$, during noise spectroscopy (Bloch sphere). 
    (c) NV and P1 level structure in the presence of a magnetic field, $B$, applied along the NV axis.
    We work within an effective spin-1/2 subspace of the NV center, $\{\ket{0}, \ket{-1}\}$, with level splitting, $\omega_\text{NV}$.
    The corresponding P1 splitting, $\omega_{P1}$, is strongly off-resonant from the NV transition. %$\Delta_0 - \omega_\text{NV} \propto B$, where $\Delta_0$ is the NV's zero-field splitting.
    (d) Secondary ion mass spectrometry (SIMS) measurement of the density of $^{14}$N as a function of depth for  sample S1. 
    The presence of a 2D layer is indicated by a sharp Nitrogen peak with a SIMS-resolution-limited $8$~nm width.
    (e) The overlap between the many-body spectral function (blue) and the power spectrum of the filter function $|f(\omega; t)|^2$  determines the variance of the phase $\sim \chi(t)$ [Eqn.~\ref{eq:chi}]. $|f(\omega; t)|^2$ for both  a Ramsey/DEER pulse sequence (purple) and a spin echo pulse sequence (orange) are shown.
    (f) Schematic depiction of the variance of the phase, $\braket{\phi^2} = -2\log C(t)$, as a function of the measurement duration $t$, for  both Ramsey/DEER (purple) and spin echo (orange).
    The labeled slopes indicate the predicted stretch powers in both the early-time ballistic regime and the late-time random-walk regime~[Table~\ref{tab:table1}]; the crossover occurs at the correlation time, $\tau_c$. }
    \label{fig:fig1}
\end{figure}

    We first outline a framework, building upon classic results in NMR spectroscopy, for understanding the decoherence dynamics of probe spins coupled to an interacting many-body system; this will enable us to present a unified theoretical background for understanding the experimental results in subsequent sections~\cite{anderson1953exchange,  klauder1962spectral, hu1974theory, salikhov1981theory, fel1996configurational, cucchietti2005decoherence, de_Sousa_2009, kubo2012statistical}.
    % Let us begin by providing a framework for understanding the decoherence dynamics of probe spins coupled to an interacting many-body system~\cite{anderson1953exchange,  klauder1962spectral, hu1974theory, salikhov1981theory, fel1996configurational, cucchietti2005decoherence, de_Sousa_2009, wang2012comparison, kubo2012statistical}.
    %
    The dynamics of a single probe spin generically depend on three properties: (i) the nature of the system-probe coupling, (ii) the system's many-body Hamiltonian $H_\textrm{int}$, and (iii) the measurement sequence itself. 
    Crucially, by averaging across the dynamics of many such probe spins, one can extract global features of the  many-body system~[Fig.~\ref{fig:fig1}(b)].
    We distinguish between two types of ensemble averaging which give rise to distinct signatures in the decoherence: (i) an average over {\it many-body trajectories} (i.e.~both  spin configurations and dynamics) yields information about the microscopic spin fluctuations 
    (for simplicity, we focus our discussion on the  infinite-temperature limit, and the analysis can be extended to finite temperature), (ii) an average over \textit{positional randomness} (i.e~random locations of the system spins) yields information about both dimensionality and  disorder.

    To be specific, let us consider a single spin-1/2 probe  coupled to a many-body ensemble via long-range, $1/r^\alpha$ Ising interactions:
    \begin{equation}\label{eq:Hz}
        H_z = \sum_i \frac{J_z}{r_i^\alpha} \hat{s}_p^z \hat{s}_i^z,
    \end{equation}
    where $r_i$ is the distance between the probe spin $\hat{s}_p$ and the $i$-th system spin $\hat{s}_i$, and Ising coupling strength $J_z$ implicitly includes any angular dependence. 
    Such power-law interactions are ubiquitous in solid-state,  atomic and molecular quantum platforms (e.g.~RKKY interactions, electric/magnetic dipolar interactions, van der Waals interactions, etc.).

     Physically, the system spins generate an effective magnetic field at the location of the probe (via Ising interactions), which can be measured with Ramsey spectroscopy~[inset, Fig. \ref{fig:fig1}(e)]~\cite{schweiger2001principles}. 
     In particular, we envision initially preparing the probe in an eigenstate of $\hat{s}_p^z$ and subsequently rotating it with a $\pi/2$-pulse such that the initial normalized coherence is unity, $C \equiv 2\braket{\hat{s}_p^x} = 1$. 
     The magnetic field, which  fluctuates due to many-body interactions,  causes the probe to Larmor precess~[inset, Fig. \ref{fig:fig1}(b)]~\cite{SM}. The phase associated with this Larmor precession  can be  read out via a population imbalance, after a  second $\pi/2$ pulse.
        
    \textit{Average over many-body trajectories}---For a many-body system at infinite temperature, $C(t) = 2\textrm{Tr}[\rho(t)\hat{s}_p^x]$, where $\rho(t)$ is the full density matrix that includes both the system and the probe. 
    The spin fluctuations are determined by the microscopic details of the many-body dynamics whose full analysis is intractable. To make progress, we approximate each spin as a stochastic classical variable $\hat{s}_i^z(t) \to s_i^z(t)$.
    The statistical properties of such variables, and their resulting ability to capture the experimental observations, provide important insights into the nature of fluctuations in strongly-interacting spin systems.

    The phase of the Larmor precession is given by $\phi (t) = \int_0^tdt'~J_z \sum_i s_i^z(t') /r_i^\alpha$.
    Assuming that $\phi(t)$ is Gaussian-distributed,
   %Assuming that $s_i^z$ are independent and Gaussian-distributed (in which case $P(\phi(t))$ is also Gaussian), 
   one finds that the average probe coherence decays exponentially as $C(t) = \braket{\text{Re}[e^{-i\phi(t)}]}=e^{-\braket{\phi^2}/2}$,  where $\braket{\phi^2} \sim \sum_i J_z^2\chi(t)/r_i^{2\alpha}$~\cite{herzog1956transient, hanson2008coherent, de2010universal,  yang:2017, SM}.  
    Here,  $\chi(t)$ encodes the response of the probe spins to the noise spectral density, $S(\omega)$, of the many-body system:
    \begin{equation}\label{eq:chi}
        \chi(t) \equiv \int d\omega~ |f(\omega; t)|^2 S(\omega),
    \end{equation}
    where $f(\omega; t)$ is the filter function associated with a particular pulse sequence (e.g.~Ramsey spectroscopy or spin echo) of total duration $t$~[Fig.~\ref{fig:fig1}(e)].

Intuitively, $S(\omega)$ quantifies the noise power density of spin flips in the many-body system; it is the Fourier transform of the autocorrelation function, $\xi(t) \equiv 4\braket{s_i^z(t) s_i^z(0)}$, and captures the spin dynamics at the level of two-point correlations~\cite{kogan2008electronic}. For Markovian dynamics, $\xi(t) = e^{-|t|/\tau_c}$, where $\tau_c$ defines the correlation time after which a  spin, on average, retains no memory of its initial orientation. In this case, $S(\omega)$ is Lorentzian and one can derive an  analytic expression for $\chi$~\cite{hu1974theory, salikhov1981theory, fel1996configurational, wang2013spin, SM}. 

A few remarks are in order. First, the premise that many-body Hamiltonian dynamics produce Gaussian-distributed phases $\phi(t)$---while oft-assumed---is challenging to analytically justify \cite{mims1968phase,klauder1962spectral,chiba1972electron,salikhov1981theory,witzel2006quantum}. Indeed, a well-known counterexample of non-Gaussian spectral diffusion occurs when the spin dynamics can be modeled as telegraph noise -- i.e. stochastic jumps between discrete values $s_i^z = \pm s_i$~\cite{chiba1972electron, abe2004electron}; the precise physical settings where such noise emerges remains the subject of active debate~\cite{mims1968phase, klauder1962spectral, chiba1972electron, abe2004electron,zhong2015optically, de2003theory, wang2013spin,bauch2020decoherence, hanson2008coherent,de2010universal}. %witzel2005quantum,
Second, we note that our Markovian assumption is not necessarily valid for a many-body system at early times or for certain forms of interactions, which can also affect the decoherence dynamics.

\textit{Average over positional randomness}---The probe's decoherence depends crucially on the spatial distribution of the spins in the many-body system. For disordered spin ensembles, explicitly averaging over their random positions yields a decoherence profile:
    \begin{equation}
        C(t)   = \int \prod_{i=1}^N \frac{d^D r_i}{V} \exp\left[\frac{- J_z^2 \chi(t)}{2 r_i^{2\alpha}} \right] = e^{- a n[J_z^2 \chi(t)]^{D/2\alpha}},
        \label{eq:intC}
    \end{equation}
    where $a$ is a dimensionless constant, $N$ is the number of system spins in a $D$-dimensional volume $V$ at a density $n \equiv N/V$~\cite{fel1996configurational, SM}. By contrast, for spins on a lattice or for a single probe spin, the exponent of the coherence scales as $\sim J_z^2 \chi(t)$~\cite{SM}.

A resonance-counting argument underlies the appearance of both the dimensionality and the interaction power-law in Eqn.~(\ref{eq:intC}). Roughly, a probe spin is only coupled to system spins that induce a phase variance larger than some cutoff $\epsilon$. This constraint on the minimum variance defines a volume of radius $r_\text{max} \sim (J_z^2 \chi(t)/\epsilon)^{1/2\alpha}$ containing   $N_s \sim n r_\text{max}^D$ spins, implying that the total variance accrued at any given time is $\epsilon N_s \sim [J_z^2 \chi(t)]^{D/2\alpha}$. Thus, the positional average simply serves to count the number of spins to which the probe is coupled.

\textit{Decoherence profile}---The functional form of the probe's decoherence, $C(t)$, encodes a number of features of the many-body system. We begin by elucidating them in the context of Ramsey spectroscopy. First, one expects a somewhat sharp cross-over in the behavior of $C(t)$ at the correlation time $\tau_c$. For early times, $t \ll \tau_c$, the phase variance accumulates as in a ballistic trajectory with $\chi \sim t^2$, while for late times, $t \gg \tau_c$, the variance accumulates as in a random walk with $\chi \sim t$~\cite{anderson1953exchange, hu1974theory, salikhov1981theory}. This leads to a simple prediction: namely, that the stretch-power, $\beta$, of the probe's exponential decay (i.e.~$-\log C(t) \sim t^\beta$)  changes from $D/\alpha$ to $D/2\alpha$ at the correlation time [Fig.~\ref{fig:fig1}(f)].

Second, moving beyond Ramsey measurements by changing the filter function, one can probe more subtle properties of the many-body noise. In particular, a spin-echo sequence filters out the leading order DC contribution from the many-body noise spectrum, allowing one to investigate higher-frequency correlations of the spin-flip dynamics. Different types of spin-flip dynamics naturally lead to different phase distributions. For the case of Gaussian noise, one finds that (at early times) $\chi \sim t^3$; however, in the case of telegraph noise the analysis is more subtle, since higher-order moments of $\phi(t)$ must be taken into account. This leads to markedly different early-time predictions for $\beta$---dependent on both the measurement sequence as well as the many-body noise~[Table~\ref{tab:table1}].

At late times, however, one  expects the probe's coherence to agree across different pulse sequences and spin-flip dynamics. For example, in the case of spin-echo, the decoupling $\pi$-pulse [inset, Fig.~\ref{fig:fig1}(e)] is ineffective on timescales larger than the correlation time,  since the spin configurations during the two halves of the free evolution are completely uncorrelated. Moreover, this same loss of correlation implies that the phase accumulation is characterized by incoherent Gaussian diffusion regardless of the specific nature of the spin dynamics (e.g. Markovian versus non-Markovian, or continuous versus telegraph).

%%%%%%%%%%%%%%%%%%%%%%%%%%%%%%%%%%%%%%%%%%
%% EXPERIMENT SUBSECTION %%
%%%%%%%%%%%%%%%%%%%%%%%%%%%%%%%%%%%%%%%%%%
\subsection*{Experimentally probing many-body noise in strongly-interacting spin ensembles}
Our experimental samples contain a high density of spin-1/2 P1 centers [blue spins, Fig.~\ref{fig:fig1}(b)], which form a strongly-interacting many-body system coupled via magnetic dipole-dipole interactions:
\begin{equation}\label{eq:Hdip}
H_\text{int} = \sum_{i<j} \frac{J_0}{r_{ij}^3} \left[ c_{ij} (\hat{s}_i^+ \hat{s}_j^- + \hat{s}_i^- \hat{s}_j^+) + \tilde{c}_{ij}\hat{s}_i^z \hat{s}_j^z \right],
\end{equation}
where $J_0 = 2\pi \times 52~$MHz$\cdot$nm$^3$, $r_{ij}$ is the distance between P1 spins $i$ and $j$, and $c, \tilde{c}$ capture the angular dependence of the dipolar interaction~\cite{SM}. We note that $H_\text{int}$ contains only the energy-conserving terms of the dipolar interaction.

The probes in our system are spin-1 NV centers, which can be optically initialized to $\ket{m_s = 0}$ using 532~nm laser light. An applied magnetic field $B$ along the NV axis  splits the $\ket{m_s = \pm 1}$ states,  allowing us to work within the effective spin-1/2 manifold $\{\ket{0}, \ket{-1}\}$.Microwave pulses at frequency $\omega_\text{NV}$ are used to perform coherent spin rotations (i.e.~for Ramsey spectroscopy or spin echo) within this manifold~[Fig.~\ref{fig:fig1}(c)].

Physically, the NV and P1 centers are also coupled via dipolar interactions. However, for a generic magnetic field strength, they are highly detuned, i.e.~$|\omega_\text{NV} - \omega_\text{P1}| \sim$~GHz, owing to the zero-field splitting of the NV center ($\Delta_0 = 2\pi \times 2.87$~GHz)~[Fig.~\ref{fig:fig1}(c)]. Since typical interaction strengths in our system are on the order of $\sim$~MHz, direct polarization exchange between an NV and P1 is strongly off-resonant. The strong suppression of spin-exchange interactions between NV and P1 centers simplifies the full magnetic dipole-dipole Hamiltonian to a system-probe Ising coupling of precisely the form given by Eqn.~\ref{eq:Hz} with  $\alpha = 3$~\cite{SM}.

%%%%%%%%%%%%%%%%%%%%%%%%%%%%%%%%%%%%%%%%%%%%%%%%%%%%%%%%
%%%%%%%%%%%%%%%%%%%%%%%%%%%%%%%%%%%%%%%%%%%%%%%%%%%%%%%%
{\it Delta-doped sample fabrication}---Sample S1 was grown via homoepitaxial plasma-enhanced chemical vapor deposition (PECVD) using isotopically purified methane (99.999$\%$ $^{12}$C)  ~\cite{ohno2012engineering}.  The delta-doped layer was formed by introducing natural-abundance nitrogen gas during growth (5~sccm, 10 minutes) in between nitrogen-free buffer and capping layers. To create the vacancies necessary for generating NV centers, the sample was electron-irradiated with a transmission electron microscope set to 145~keV~\cite{mclellan2016patterned} and subsequently annealed at 850$^{\circ}~$C for 6 hours.

\begin{figure}[H]
    \centering
    \includegraphics[width=3.2in]{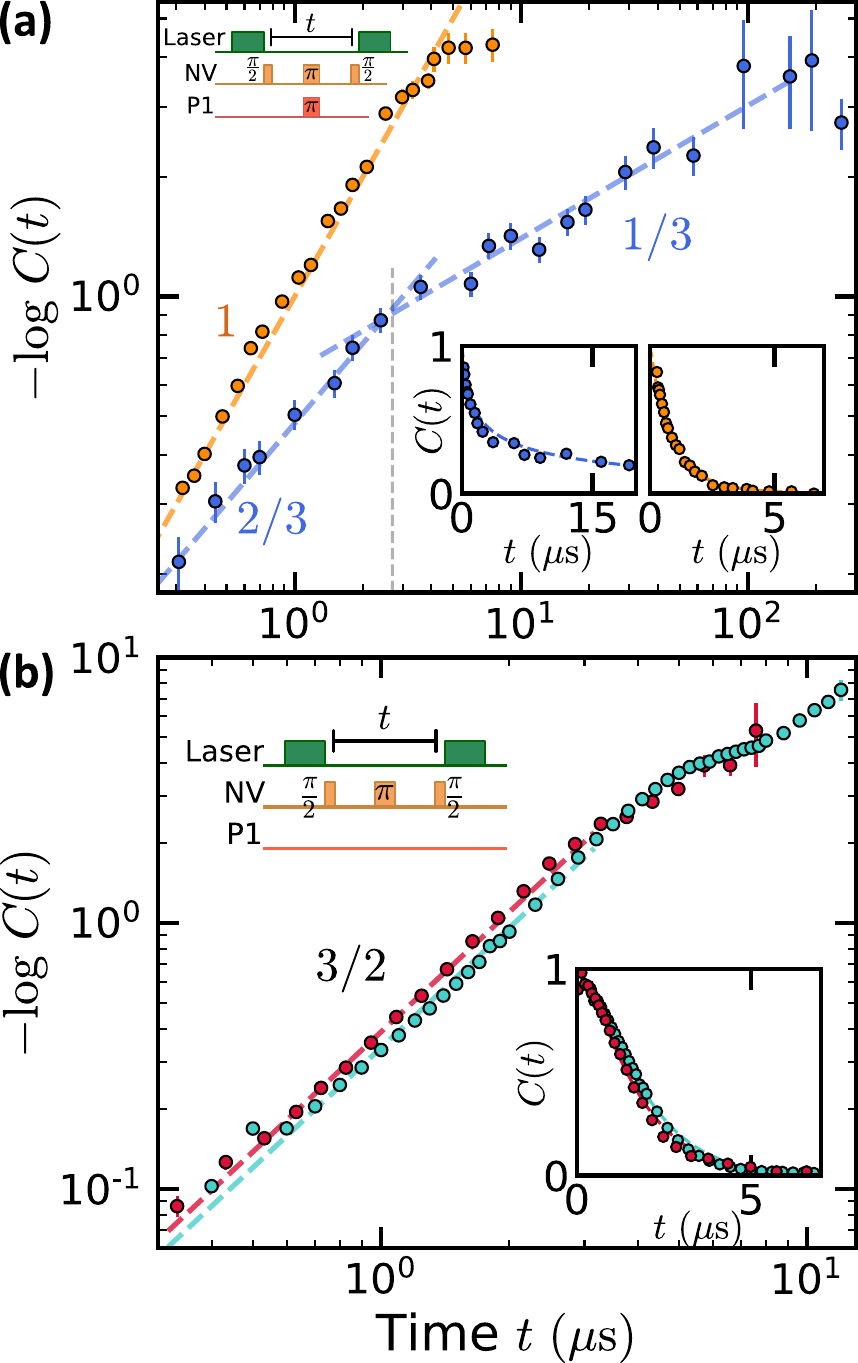}
\caption{
    \textbf{Dimensionality and dynamics from many-body noise} (a) Depicts the normalized coherence for a DEER measurement on sample S1 (blue) and sample S2 (yellow) as a function of the free evolution time $t$. Dashed blue lines indicate the predicted early- and late-time stretch powers of 2/3 and 1/3, respectively, for a dipolar spin system in two dimensions. 
    Dashed yellow line depicts the predicted early-time stretch power of 1 for a dipolar spin system in three dimensions~[Table~\ref{tab:table1}].
     Together, these data demonstrate the two- and three-dimensional nature of samples S1 and S2, respectively.
     Lower right insets show the same data on a linear scale.
     Top left inset shows the DEER pulse sequence.
     (b)
    Spin echo measurements on three-dimensional dipolar spin ensembles in samples S3 (teal) and S4 (pink) clearly exhibit a stretch power of 3/2 (dotted lines) over nearly two decades in time. 
    This is consistent with the presence of Gaussian noise and allows one to explicitly rule out telegraph noise. 
    Lower right inset shows the same data on a linear scale. 
    Top left inset shows the spin echo pulse sequence. All data are presented as mean values $\pm$ SEM. %statistical error from photon shot noise.}
    }
    \label{fig:fig2}
\end{figure}

%%%%%%%%%%%%%%%%%%%%%%%%%%%%%%%%%%%%%%%%%%%%%%%%%%%%%%%%
%%%%%%%%%%%%%%%%%%%%%%%%%%%%%%%%%%%%%%%%%%%%%%%%%%%%%%%%
{\it Two-dimensional spin dynamics}---We begin by performing double electron-electron resonance (DEER) measurements on sample S1. While largely analogous to Ramsey spectroscopy [Table~\ref{tab:table1}], DEER has the technical advantage that it filters out undesired quasi-static fields (e.g.~from hyperfine interactions between the NV and host nitrogen nucleus)~\cite{schweiger2001principles, eichhorn2019optimizing}. As shown in  Fig.~\ref{fig:fig2}(a) [blue data, inset], the NV's coherence decays on a time scale $\sim 5~\mu$s.

To explore the functional form of the probe NV's decoherence, we plot the negative logarithm of the coherence, $-\log C(t)$, on a log-log scale, such that the stretch power, $\beta$, is simply given by the slope of the data.  At early times, the data exhibit $\beta = 2/3$ for over a decade in time [blue data, Fig.~\ref{fig:fig2}(a)]. At a timescale $\sim 3 ~\mu$s (vertical dashed line), the data crosses over to a stretch power of $\beta = 1/3$ for another decade in time. This behavior is in excellent agreement with that expected for two-dimensional spin dynamics driven by dipolar interactions [Fig.~\ref{fig:fig1}(f), Table~\ref{tab:table1}].

For comparison, we perform DEER spectroscopy on a conventional three-dimensional NV-P1 system (sample S2, see Methods). 
As shown in Fig.~\ref{fig:fig2}(a) (orange), the data exhibit $\beta = 1$ for a decade in time, consistent with the prediction for three-dimensional dipolar interactions [Table~\ref{tab:table1}].
However, the crossover to the late-time ``random walk'' regime is difficult to experimentally access because the larger early-time stretch power causes a faster decay to the noise floor.

%%%%%%%%%%%%%%%%%%%%%%%%%%%%%%%%%%%%%%%%%%%%
%%%%%%%%%%%%%%%%%%%%%%%%%%%%%%%%%%%%%%%%%%%%
{\it Characterizing microscopic spin-flip dynamics}---To probe the nature of the microscopic spin-flip dynamics in our system, we perform spin-echo measurements on  three dimensional samples [S3, S4 (Type IB)], which exhibit a significantly higher P1-to-NV density ratio (see Methods).
% than??
For lower relative densities (i.e.~samples S1 and S2), the spin echo measurement contains a confounding signal from interactions between the NVs themselves (see Methods). 

In both samples (S3, S4), we find that the coherence exhibits a stretched exponential decay with $\beta = 3/2$ for well over a decade in time [Fig.~\ref{fig:fig2}(b)].  Curiously, this is consistent with  Gaussian spectral diffusion where $\beta = 3D/2\alpha = 3/2$ and patently inconsistent with  the telegraph noise prediction of $\beta = 1+ D/\alpha = 2$. 
While in agreement with prior measurements on similar samples \cite{bauch2020decoherence}, this observation is actually rather puzzling and related to a question in the context of dipolar spin noise~\cite{anderson1953exchange, herzog1956transient, klauder1962spectral, mims1968phase, fel1996configurational, kubo2012statistical, salikhov1981theory, chiba1972electron, schweiger2001principles,choi2017depolarization, zhidomirov1969contribution, glasbeek1981phase, abe2004electron, zhong2015optically, de2003theory, wang2013spin, bauch2020decoherence, hanson2008coherent,de2010universal}. %witzel2005quantum, 
In particular, one naively expects that spins in a strongly interacting system
should be treated as stochastic binary variables, thereby generating telegraph noise; for the specific case of dipolar spin ensembles, this expectation dates back to seminal work from Klauder and Anderson~\cite{klauder1962spectral}. The intuition behind this noise model is most easily seen in the language of the master equation---each individual spin ``sees'' the remaining system as a Markovian bath. The resulting local spin dynamics are then characterized by a series of stochastic quantum jumps that flip the spin orientation and give rise to telegraph noise~\cite{SM}. Alternatively, in the Heisenberg picture, the same intuition can be understood from the spreading of the  operator $\hat{s}_i^z$; this spreading hides local coherences in many-body correlations, leading to an ensemble of telegraph-like, classical trajectories~\cite{SM}.

We conjecture that the observation of Gaussian spectral diffusion in our system is related to the presence of disorder, which strongly suppresses operator spreading~\cite{witzel2012quantum}. 
To illustrate this point, consider the limiting case where the operator dynamics are constrained to a single spin. In this situation, the  dynamics of $\hat{s}_i^z(t)$ follow a particular coherent trajectory around the Bloch sphere, and the rate at which the probe  accumulates phase is continuous~\cite{SM}. Averaging over different trajectories of the coherent dynamics naturally leads to Gaussian noise~\cite{SM}.

\begin{figure}[H]
    \centering
    \includegraphics[width=6in]{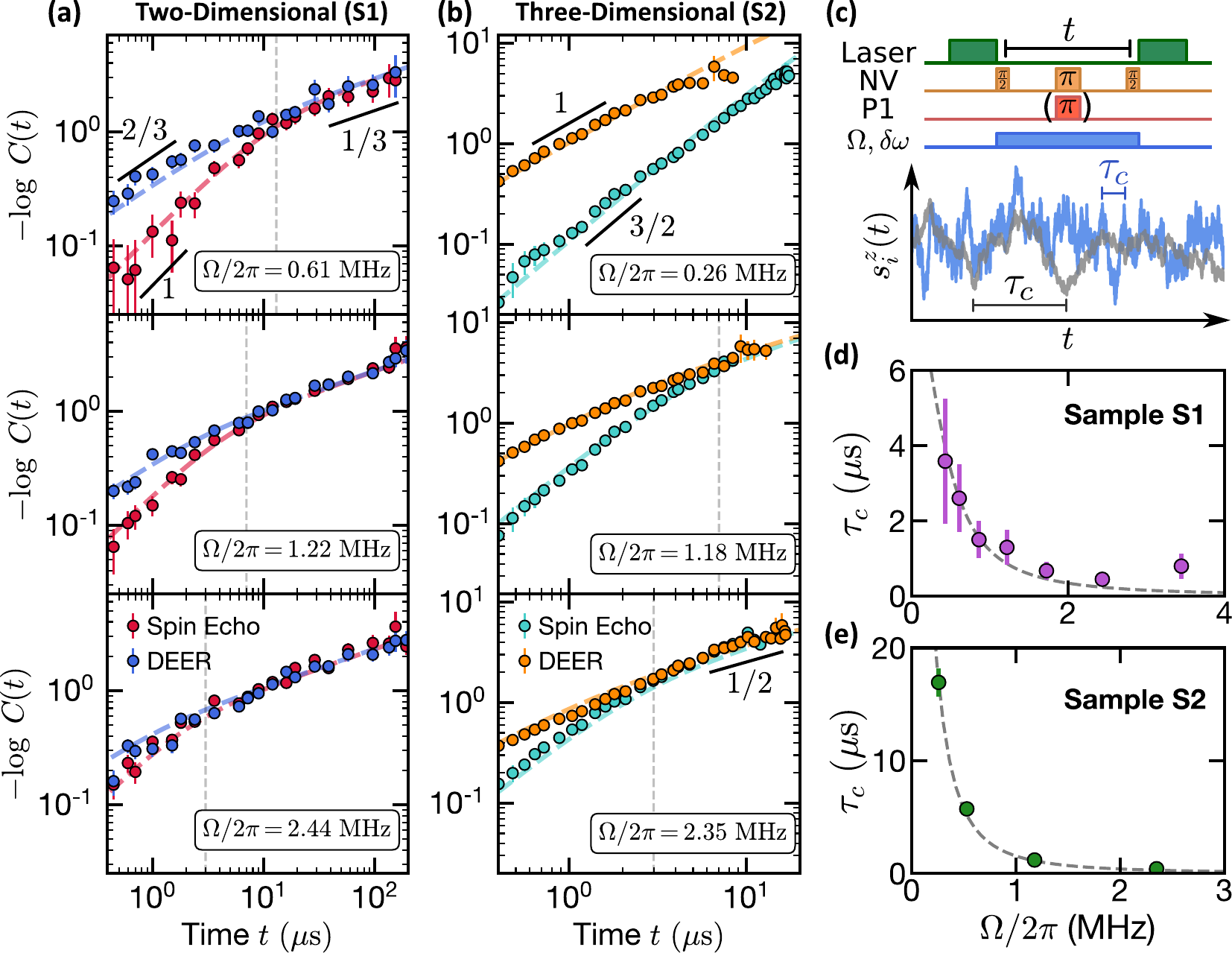}
\caption{ \textbf{Tuning the correlation time of the bath}
(a,b) Measurements of DEER (blue, orange) and spin echo (red, teal) on two- and three-dimensional samples (S1, S2) for different powers of the polychromatic (i.e.~incoherent) drive at fixed linewidth $\delta \omega = 2\pi \times (18, 20)~$MHz, respectively. The time at which the two signals overlap  (vertical dashed lines)  functions as a proxy for the correlation time and decreases as the power of the incoherent driving increases (top to bottom panels). The data is well-fit by analytic expressions for $[\chi(t)]^{D/2\alpha}$ [Eqn.~\ref{eq:chi_deer}] (dashed curves). Data are presented as mean values $\pm$ SEM. (c) An incoherent drive field (light blue) with power $\sim \Omega^2$ and linewidth $\delta \omega$ is applied to the P1 spins during the free evolution time $t$ of both DEER and spin echo sequences in order to tune the correlation time of the many-body system. In this case, $s_i^z(t)$ evolves as a Gaussian random process schematic for short and long correlation time $\tau_c$). (d,e) The correlation times, $\tau_c$, extracted from fitting the data to Eqn.~\ref{eq:chi_deer}  for samples S1 (purple) and S2 (green) are plotted as a function of $\Omega$, and agree well with a simple theoretical model (dashed grey curves)~\cite{SM} Data are presented as best fit values $\pm$ fitting error. }
    \label{fig:fig3}

\end{figure}

\emph{Controlling the P1 spectral function}---Next, we demonstrate the ability to directly control the P1 noise spectrum for both two- and three-dimensional dipolar spin ensembles (i.e.~samples S1, S2). In particular, we engineer the shape and linewidth of $S(\omega)$ by driving the P1 system with a polychromatic microwave tone~\cite{Ernst1966}.
This drive is generated by adding phase noise to the resonant microwave signal at $\omega_\textrm{P1}$ in order to produce a Lorentzian drive spectrum with linewidth $\delta \omega$~[Fig.~\ref{fig:fig3}(c)].
While such techniques originated in the context of  broadband noise decoupling~\cite{Ernst1966},  here, we directly tune the correlation time of the P1 system and measure a corresponding change in the crossover timescale between coherent and incoherent spin dynamics~\cite{glasbeek1981phase, salikhov1981theory}.

Microscopically, the polychromatic drive leads to a number of physical effects.
First, tuning the Rabi frequency, $\Omega$, of the drive provides a direct knob for controlling the correlation time, $\tau_c$, of the P1 system. 
Second, since the many-body system inherits the noise spectrum of the drive, one has provably Gaussian statistics for the spin variables $s_i^z$~\cite{SM}. 
Third, our earlier Markovian assumption is explicitly enforced by the presence of a Lorentzian noise spectrum.
Taking these last two points together allows one to analytically predict the precise form of the NV probe's decoherence profile, $-\log C(t) \sim \chi(t)^{D/2 \alpha}$, for either DEER or spin-echo spectroscopy:
\begin{equation}
% \begin{align}
\begin{split}
\label{eq:chi_deer}
&\chi^{\textrm{DEER}} (t) = 2\tau_ct-2\tau_c^2\left(1-\mathrm{e}^{-\frac{t}{\tau_c}}\right), \\ 
&\chi^{\textrm{SE}} (t) = 2\tau_ct-2\tau_c^2 \left(3+\mathrm{e}^{-\frac{t}{\tau_c}}-4\mathrm{e}^{-\frac{t}{2\tau_c}}\right).
\end{split}
\end{equation}

We perform both DEER and spin-echo measurements as a function of the power ($\sim \Omega^2$) of the polychromatic drive for our two-dimensional sample (S1) [Fig.~\ref{fig:fig3}(a)]. 
As expected, for weak driving [top, Fig.~\ref{fig:fig3}(a)], the DEER signal (blue) is analogous to the undriven case, exhibiting a cross-over from a stretch power of $\beta = 2/3$ at early times to  a stretch power of $\beta = 1/3$ at late times. 
For the same drive strength, the spin echo data (red) also exhibit a cross over between two distinct stretch powers, with the key difference being that $\beta = 3D/2\alpha = 1$ at early times. 
This represents an independent (spin-echo-based) confirmation of the two-dimensional nature of our delta-doped sample.

Recall that at late times (i.e.~$t \gtrsim \tau_c$), one expects the NV's coherence $C(t)$ to agree across different pulses sequences [Fig.~\ref{fig:fig1}(f)].
This is indeed borne out by the data [Fig.~\ref{fig:fig3}]. 
In fact, the location of this late-time overlap provides a proxy for estimating the correlation time and is shown as the dashed grey lines in Fig.~\ref{fig:fig3}(a). 
As one increases the power of the drive [Fig.~\ref{fig:fig3}(a)], the noise spectrum, $S(\omega)$, naturally broadens.
In the data, this manifests as a shortened correlation time, with the location of the DEER/echo overlap shifting to earlier time-scales  [Fig.~\ref{fig:fig3}(a)].

Analogous measurements on a three-dimensional spin ensemble (sample S2), reveal much the same physics [Fig.~\ref{fig:fig3}(b)], with stretch powers again consistent with a Gauss-Markov prediction [Table~\ref{tab:table1}]. 
For weak driving, $C(t)$ is consistent with the early-time ballistic regime for over a decade in time~[Fig.~\ref{fig:fig3}b, top panel]; however, it is difficult to access late enough time-scales to observe an overlap between DEER and spin echo. 
Crucially, by using the drive to push to shorter correlation times, we can directly observe  the late-time random-walk regime in three dimensions, where $\beta = 1/2$~[Fig.~\ref{fig:fig3}b, middle and bottom panels].

Remarkably, as evidenced by the dashed curves in Fig.~\ref{fig:fig3}(a,b), our data exhibit excellent agreement---across different dimensionalities, drive strengths, and pulse sequences---with the analytic predictions presented in Eqn.~\ref{eq:chi_deer}.
Moreover, by fitting  $\chi^{D/2\alpha}$ simultaneously across spin echo and DEER datasets for each $\Omega$, we quantitatively extract the correlation time, $\tau_c$. 
Up to an $\mathcal{O}(1)$ scaling factor, we find that the extracted $\tau_c$ agrees well with the DEER/echo overlap time.
In addition, the behavior of  $\tau_c$ as a function of $\Omega$ also exhibits quantitative agreement with an analytic model that predicts $\tau_c \sim \delta\omega/\Omega^2$ in the limit of strong driving [Fig.~\ref{fig:fig3}(d,e)]~\cite{SM}.   %and thus $\tau_c$, changing qualitatively as we expect: the time

We emphasize that although one observes $\beta = 3D/2\alpha$  in both the driven [Fig.~\ref{fig:fig3}(a,b)] and undriven [Fig.~\ref{fig:fig2}(b)] spin echo measurements, the underlying physics is extremely different. %
In the latter case, Gaussian spectral diffusion emerges from isolated, disordered, many-body dynamics, while in the former case, it is imposed by the external drive.

\subsection*{A two-dimensional solid-state platform for quantum simulation and sensing}\label{sec:platform}
%\subsection*{Towards Quantum simulation and sensing with P1 centers}
%
Our platform offers two distinct paths toward quantum simulation and sensing using strongly-interacting, two-dimensional, spin-polarized ensembles. 
%
%dominant dynamics driven by interactions among themselves?
%
%
%First, treating the NV centers themselves as the many-body system enables straightforward optical polarization of the spins, but it is not clear whether interactions in this relatively dilute ensemble dominate compared to disorder.
%
First, treating the NV centers themselves as the many-body system directly leverages their optical polarizability. 
However, given their relative diluteness, it is natural to ask whether one can access regimes where the NV-NV interactions dominate over other energy scales.
Conversely, treating the P1 centers as the many-body system takes advantage of their higher densities and interaction strengths, with the key challenge being that these dark spins cannot be optically pumped.
%
%After decoupling the NV-P1 interactions, we show that NV-NV interactions are the dominant contribution to the dynamics. 
%
Here, we demonstrate that both of these paths are viable for sample S1:
(i) we show that the dipolar interactions among NV centers can dominate their decoherence dynamics, using advanced dynamical decoupling sequences; (ii) we demonstrate direct polarization exchange between NV and P1 centers, providing a mechanism to spin-polarize the P1 system.

\begin{figure}[H]
    \centering
    \vspace{-2.5cm}
    \includegraphics[width=2.5in]{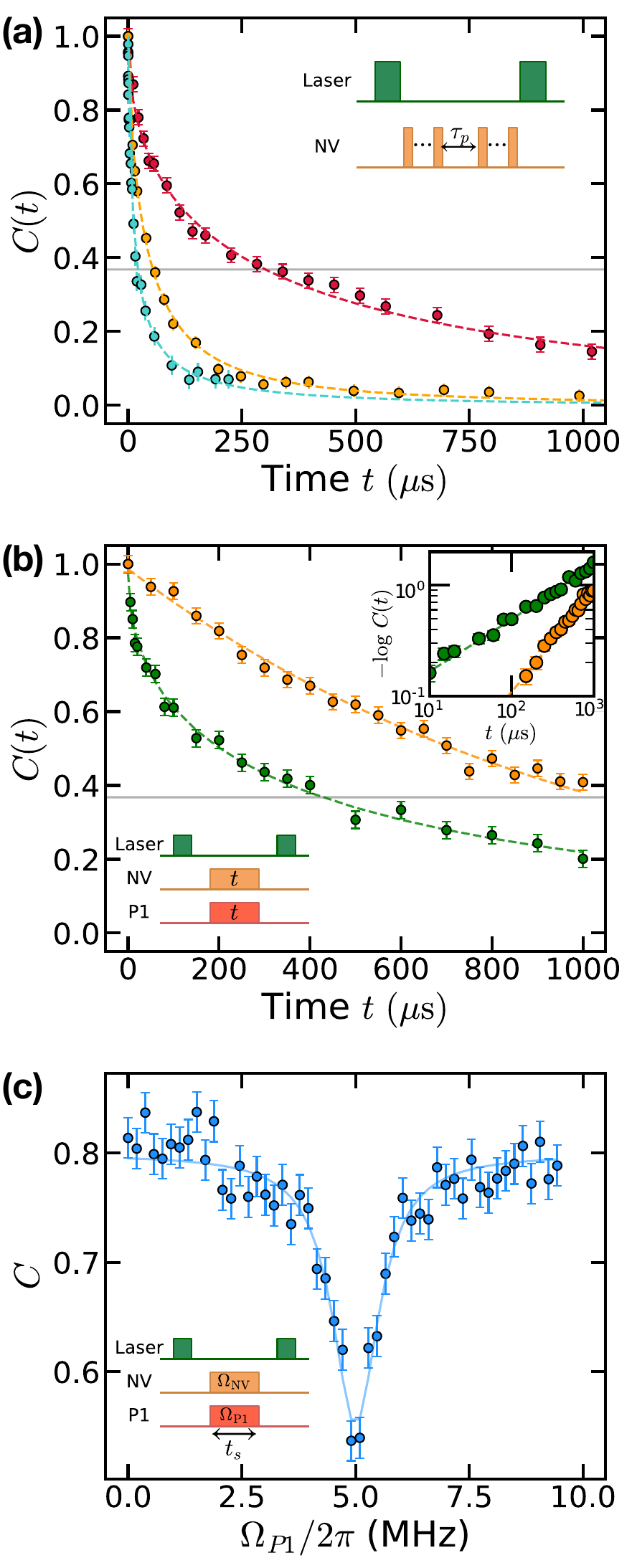}
    \caption{
    \textbf{Hybrid two-dimensional spin system for simulation and sensing} (a) We measure $T_2$ with spin echo (blue), XY-8 (orange) and DROID (red) pulse sequences with interpulse spacing $\tau_p = 100$~ns. The spin echo and XY-8 decoherence profiles are fitted by Eqn.~\ref{eq:chi_deer} while DROID is fitted by a stretched exponential $\sim e^{-(t/T_2)^{0.5}}$. The $1/e$ lifetimes (gray line) for spin echo, XY-8 and DROID are 22~$\mu$s, 47~$\mu$s and 302~$\mu$s, respectively.  Inset: schematic of dynamical decoupling sequence with interpulse spacing $\tau_p$. 
(b) Spin-locked NV depolarization profile. Only NV centers (orange) or both NV and P1 centers (green) are driven at a Rabi frequency of $2\pi\times 5$ MHz. The resonant spin-exchange interactions reduce the spin-locking relaxation time $T_{1\rho}$ by a factor of $\sim 3$. Top inset shows the same data plotted on a log-log scale to elucidate stretch powers. Bottom inset: spin-locking pulse sequence. 
(c) Hartmann-Hahn polarization exchange resonance. The NV Rabi frequency $\Omega_\text{NV}$ is fixed at $2\pi\times5$ MHz. When the P1 Rabi frequency $\Omega_\text{P1}$ matches $\Omega_\text{NV}$, a reduction in contrast is induced by the resonant polarization exchange between NV and P1 centers. The data is fitted by a Lorentzian (dashed curve) with a linewidth of $2\pi\times1.2$ MHz. Inset: Hartmann-Hahn pulse sequence, with fixed spin-locking duration $t_s=200~\mu$s. All data are presented as mean values $\pm$ SEM.
}
    \label{fig:fig4}

\end{figure}

\emph{Interacting NV ensemble---}To demonstrate NV-NV-interaction-dominated dynamics, we compare the decoherence timescales between spin echo, XY-8, and DROID dynamical decoupling sequences~\cite{ZhouQuantumMetrologyStrongly2020}.
%sequence H of this paper
%
%
The spin echo effectively decouples static disorder, while the XY-8 sequence further decouples NV-P1 interactions.
As depicted in Fig.~\ref{fig:fig4}(a),  XY-8 pulses extend the spin-echo decay time (defined as the $1/e$-time) by approximately a factor of two. 
With NV-P1 interactions decoupled, our hypothesis is that the dynamics are now driven by dipolar interactions \emph{between} the NV centers.
To test this, we perform a DROID decoupling sequence, which eliminates the dipolar dynamics between NV centers~\cite{ZhouQuantumMetrologyStrongly2020} (see Methods).
Remarkably, this extends the coherence time by nearly an order of magnitude, demonstrating that NV-NV interactions are, by far, the dominant source of many-body dynamics in this regime. 
Moreover, the XY-8 decoherence thus provides an estimate of an average NV spin-spin spacing of 15~nm. %provides an estimate of the average NV-NV interaction strength in our system, $\sim 2\pi\times15~$kHz, consistent with a 15~nm spin-spin spacing (see Methods).

%Indeed, the DROID sequence extends the XY-8 $1/e$-time by nearly an order of magnitude, which may be limited only by technical constraints on the interpulse spacing $\tau_p \geq 100~$ns~[Fig.~\ref{fig:fig4}(a)]. 

%NV centers not non-interacting ensemble

\emph{Interacting P1 ensemble---}The polarization of the optically-dark P1 ensemble can be realized by either (i) working at low temperatures and large magnetic fields~\cite{takahashi2008quenching}, or (ii) using NV centers to transfer polarization to the P1 centers. Here, we focus on the latter. 
While NV-P1 polarization transfer has previously been  demonstrated~\cite{BelthangadyDressedStateResonantCoupling2013, LaraouiApproachDarkSpin2013}, it has not been measured in a two-dimensional system; indeed, conjectures about  localization in such systems indicate that polarization transfer could be highly suppressed~\cite{burin:2015,yao:2014}. 

To investigate, we employ a Hartmann-Hahn sequence designed to transfer polarization between NV and P1 spins in the rotating frame~\cite{BelthangadyDressedStateResonantCoupling2013, LaraouiApproachDarkSpin2013}.  
%
%We drive the NV and P1 centers at Rabi frequencies $\Omega_\text{NV} = 2\pi\times 5~$MHz and $\Omega_\text{P1}$, respectively.
%
In particular, we drive the NV and P1 spins independently, with Rabi frequencies $\Omega_\text{NV}$ and $\Omega_\text{P1}$. 
When only the NV centers are driven, we are effectively performing a so-called spin-locking measurement~\cite{hartmann1962nuclear}; for $\Omega_\text{NV} = 2\pi \times 5~$MHz, we find that the NV centers depolarize on a timescale $T_{1\rho} = 1.05(3)~$ms~[Fig.~\ref{fig:fig4}b, orange]. 
The data are cleanly fit by a simple exponential and consistent with phonon-limited decay [Fig.~\ref{fig:fig4}b, inset]. 
By contrast, when the driving satisfies the Hartmann-Hahn condition, $\Omega_\text{NV} = \Omega_\text{P1}$, the NV and P1 spins can resonantly exchange polarization. 
To  characterize this,  we fix $\Omega_\text{NV} = 2\pi\times 5~$MHz and choose a  spin-locking duration $t_s =200~\mu$s.
By sweeping the  P1 Rabi frequency,  we indeed observe a resonant polarization-exchange feature centered at  $\Omega_\text{P1} = 2\pi\times 5~$MHz with a linewidth $\sim 2\pi\times1.2~$MHz~[Fig.~\ref{fig:fig4}(c)], consistent with the  intrinsic P1 linewidth.
%
% This is indeed borne out by the data.
%
As illustrated in Figure~\ref{fig:fig4}(b), on resonance, the NV depolarization is significantly enhanced via polarization transfer to the P1 centers and the data exhibit a three-fold decrease in the decay time.
Moreover, the data are well-fit with a stretch power $\beta = 1/3$ [Fig.~\ref{fig:fig4}b, inset], which is also indicative of interaction-dominated decay~\cite{choi2017depolarization} (see Methods).

\subsection*{Conclusion and Outlook}
Our results demonstrate the diversity of information that can be accessed via the decoherence dynamics of a probe spin ensemble.
For example, we shed light on a long-standing debate about the nature of spin-flip noise in a strongly-interacting dipolar system~\cite{mims1968phase, klauder1962spectral, chiba1972electron,zhidomirov1969contribution, glasbeek1981phase, abe2004electron, zhong2015optically, de2003theory, wang2013spin, bauch2020decoherence, hanson2008coherent,de2010universal}.
%witzel2005quantum, 
Moreover, we directly measure the correlation time of the many-body system and introduce a technique to probe its dimensionality.
% This technique solves a long-standing challenge for spin ensembles embedded in solids , 
This technique is particularly useful for disordered spin ensembles embedded in solids~\cite{cappellaro2007dynamics, lukin2020integrated}, where a direct, non-destructive measurement of nanoscale spatial properties is challenging with conventional toolsets. 
% [everyone is doing fabrication, defects in Si, defects in SiC, defects in diamond, rare earth defects in oxides, Jeff T, Andrei Faraon... ]

    One can imagine generalizing our work in a number of promising directions. 
    %
    %Our work opens the door to a number of intriguing directions.
    %
    First, the ability to fabricate and characterize strongly-interacting, two-dimensional dipolar spin ensembles opens the door to a number of intriguing questions within the landscape of quantum simulation. 
    Indeed, dipolar interactions in 2D are  quite special from the perspective of localization, allowing one to experimentally probe the role of many-body resonances~\cite{burin:2015,yao:2014}.
    In the context of ground state physics, the long-range, anisotropic nature of the dipolar interaction has also been predicted to stabilize a number of exotic phases, ranging from supersolids to spin liquids~\cite{yao2018quantum,chomaz2019long}. % 
    Connecting this latter point back to noise spectroscopy, one could imagine tailoring the probe's filter function to distinguish between different types of ground-state order.%~\cite{chatterjee2019diagnosing,khoo2021universal}. 
    
    Second, dense ensembles of two dimensional spins also promise a number of unique advantages with respect to quantum sensing~\cite{ohno2012engineering,sushkov2014magnetic, eichhorn2019optimizing}.  
    For example, a 2D layer of NVs fabricated near the diamond surface would exhibit a significant enhancement in spatial resolution (set by the depth of the layer) compared to a three-dimensional ensemble at the same density, $\rho$~\cite{ohno2012engineering,rosskopf2014investigation}. %,pham2016nmr
    In addition,  for samples where the coherence time is limited by spin-spin interactions, a lower dimensionality reduces the coordination number and leads to an enhanced $T_2$ scaling as  $n^{-\alpha/D}$~\cite{SM}. 
    %add soonwon paper

Third, one can  probe the relationship between operator spreading and Gauss-Markov noise by exploring samples with different  relaxation rates, interaction power-laws, disorder strengths and spin densities   \cite{mims1968phase,glasbeek1981phase}. One could also utilize alternate pulse sequences, such as stimulated echo, to provide a more fine-grained characterization of the many-body noise (e.g.~the entire spectral diffusion kernel)~\cite{anderson1953exchange, mims1968phase}.

Finally, our framework  can also be applied to long-range-interacting systems of Rydberg atoms, trapped ions, and polar molecules. In such systems, the ability to perform imaging and quantum control at the single-particle level allows for greater freedom in designing methods to probe many-body noise. As a particularly intriguing example, one could imagine a non-destructive, time-resolved generalization of many-body noise spectroscopy, where one 
repeatedly interrogates the probe without projecting  the many-body system. 
    
% {\it Note added:} During the completion of this work, we became aware of complementary work using single NV centers to measure the noise from a 2D disordered spin ensemble on the diamond surface~\cite{dwyer2021probing}. Both works use the decoherence profile of the probe spin to characterize the dimensionality and dynamics of the many-body system.
%{\it Note added:} During the completion of this work, we became aware of complementary work using single NV centers to measure the noise from a 2D disordered spin ensemble on the diamond surface~\cite{dwyer2021probing}, which will appear in the same arXiv posting. Both works use the decoherence profile of the probe spin to characterize the dimensionality and dynamics of the many-body system.

% \begin{acknowledgments}
\section*{Acknowledgments}
We gratefully acknowledge the insights of and discussions with M. Aidelsburger, D. Awschalom, B. Dwyer, C. Laumann, J. Moore, E. Urbach, and H. Zhou.
This work was support by the Center for Novel Pathways to Quantum Coherence in Materials, an Energy Frontier Research Center funded by the U.S. Department of Energy, Office of Science, Basic Energy Sciences (materials growth, sample characterization, and noise spectroscopy), the US Department of Energy (BES grant No. DE-SC0019241) for driving studies,  and the Army Research Office through the MURI program (grant number W911NF-20-1-0136) for theoretical studies, the W.~M.~Keck foundation,  the David and Lucile Packard foundation, and the A.~P.~Sloan foundation. E.J.D. acknowledges support from the Miller Institute for Basic Research in Science. S.A.M. acknowledges the support of the Natural Sciences and Engineering Research Council of Canada (NSERC), (funding reference number AID 516704-2018) and the NSF Quantum Foundry through Q-AMASE-i program award DMR-1906325.
D.B. acknowledges support from the NSF Graduate Research Fellowship Program (grant DGE1745303) and The Fannie and John Hertz Foundation.
% \end{acknowledgments}

\section*{Author Contributions Statement}
E.J.D., W.W., T.M, W.S., M.J., Y.L., Z.W, and C.Z.  performed the experiments. B.Y., F.M., B.K., and S.C. developed the theoretical models and methodology. E.J.D, F.M., and W.W. performed the data analysis. S.M. and A.B.J. prepared and provided the diamond substrates. A.B.J and N.Y.Y. supervised the project. E.J.D, B.Y., F.M., C.Z., N.Y.Y wrote the manuscript with input from all authors.

%%%%%%%%%%%%%%%%%%%%%%%%%%%%%%%%%%%%%%%%%%%%%%%%%%%%%%%%%%%%%%%%%%%%%%%%%%%%%%%%%%%%%%%%%%%%%%%%%%%%%%%%%%%%%%%%%%%%%%%%%%%%%%%%%%%%%%%%%%%%%%%%%%%%%%
\section*{Competing Interests Statement}
The authors declare no competing interests.
%%%%%%%%%%%%%%%%%%%%%%%%%%%%%%%%%%%%%%%%%%%%%%%%%%%%%%%%%%%%%%%%%%%%%%%%%%%%%%%%%%%%%%%%%%%%%%%%%%%%%%%%%%%%%%%%%%%%%%%%%%%%%%%%%%%%%%%%%%%%%%%%%%%%%%
\newpage
\renewcommand{\arraystretch}{2}
\setlength{\tabcolsep}{6pt}
\begin{table*}[h]
    \centerfloat
    \begin{tabular}{c|c|c|c}  \hline \hline 
         \makecell{Many-body noise properties} & \makecell{Measurement\\sequence} &  \makecell{Early-time\\ (ballistic regime)\\stretch power} & \makecell{Late-time\\(random walk regime)\\stretch power}  \\ \hline \hline 
         
         \multirow{2}{*}{
        %  Gauss-Markov Noise\\
         \makecell[c]{\includegraphics[scale=.45]{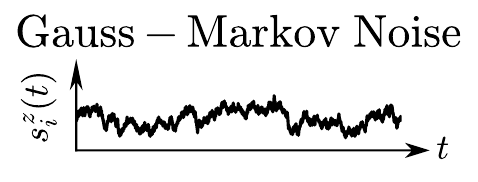}}
        %  \makecell[c]{}
         } & DEER/Ramsey & $D/\alpha$ & $D/2\alpha$ \\ \cline{2-4}
         &Spin Echo & $3D/2\alpha$ & $D/2\alpha$ \\ \hline \hline 
         \multirow{2}{*}{
         \makecell[c]{
         \includegraphics[scale=.45]{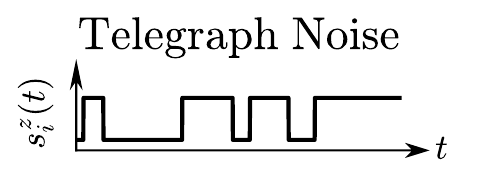}} 
        %  \makecell[c]{Telegraph Noise}
         }
         & DEER/Ramsey &  $D/\alpha$ & $D/2\alpha$  \\ \cline{2-4}
         &Spin Echo & $1 + D/\alpha$ & $D/2\alpha$  \\ \hline \hline 
    \end{tabular}
    \caption{Predicted early and late-time stretch powers of the probe spin decoherence profile when coupled to a $D-$dimensional system via power-law Ising interactions $\sim 1/r^\alpha$. We distinguish between Gaussian and telegraph spin-flip noise in the many-body system, which gives rise to different predictions for the early-time spin echo stretch power.}
    \label{tab:table1}
\end{table*}
%%%%%%%%%%%%%%%%%%%%%%%%%%%%%%%%%%%%%%%%%%%%%%%%%%%%%%%%%%%%%%%%%%%%%%%%%%%%%%%%%%%%%%%%%%%%%%%%%%%%%%%%%%%%%%%%%%%%%%%%%%%%%%%%%%%%%%%%%%%%%%%%%%%%%%
\newpage

\newpage

\newpage

\newpage

\newpage
% \section*{References}
\bibliographystyle{naturemag}
\bibliography{references}

\newpage
\section*{Methods}

% \section{Methods}
\section{Sample Preparation and Characterization}
\subsection{Sample S1}
\subsubsection{Sample fabrication}
Here, we add to the details provided in section \textit{Delta-doped sample fabrication} of the main text: sample S1 was grown on a commercially-available Element-6 electronic grade (100) substrate, polished by Syntek~\citeMethods{syntek} to a surface roughness less than 200~pm. 
%We perform homoepitaxial diamond growth of sample S1 using a SEKI AX6300 plasma-enhanced chemical vapor deposition tool~\cite{ohno2012}. 
Throughout the PECVD growth process~\cite{ohno2012engineering}, we used 400~sccm of hydrogen gas with a background pressure of 25~Torr, and a  microwave power of 750~W. The sample temperature was held at 800$^{\circ}$C.

\subsubsection{NV density}\label{sec:NVdensity}
We estimate the NV areal density in sample S1 via the XY-8 decoherence profile~\citeMethods{choi2020robust}, which is dominated by intragroup NV interactions (i.e., within the NV group aligned with the applied magnetic field $B$)~[Fig.~4(a) in main text]. 
%In particular, decoherence caused by P1 spin-flip noise is largely suppressed due to our choice of interpulse spacing $\tau_p = 100~$ns, which is much shorter than the correlation time $\tau_c$ of the P1 system. 
We therefore treat the XY-8 data as a Ramsey measurement of the average NV-NV coupling, which we convert to a density using the dipolar interaction strength $J_0 = 2\pi \times 52~$MHz${}\cdot{}$nm$^{3}$. %These NV centers, comprising one-quarter of the total NV density, cannot be decoupled from each other with $pi$ pulses. %
% For a known dipolar coupling strength $J_0 = (2\pi)\times 52~$MHz$\cdot$nm$^{3}$, the NV density $n_\text{3D}^\text{NV}$ %(with units $[n_\text{3D}^\text{NV}]~=~$nm$^{-3}$, and not to be confused with the $D-$dimensional density $n$ defined in the main text) 
% fixes an average interaction strength. To calculate the decay profile, we treat the decoherence as arising only from intragroup Ising interactions, which is correct at least at short times when the NV centers are spin-polarized.
%
We compare the XY-8 data with numerically-computed Ramsey decoherence,  which we calculate as follows: we consider a central probe NV interacting with a bath of other NVs, placed randomly in a thin slab of thickness $w$ with density $n_\text{3D}^\text{NV}/4$ (one NV group). %With dipolar coupling strength $J_0 = (2\pi)\times 52~$MHz$\cdot$nm$^{3}$, the NV density fixes an average interaction strength. 
After selecting a random spin configuration for the bath NVs, we compute the Ramsey signal $\sim \cos(\phi)$ for the probe NV. We then average over many such samples, and the resulting curve %corresponds to an average of the coherent NV oscillations over the positional and configurational realizations of the bath NVs, and 
exhibits a stretched exponential decay of the form $C(t) = e^{-(t/T_2)^{2/3}}$. This functional form matches our expectation for the early-time ballistic regime~[Table~1 in main text], because we have not included flip-flop dynamics in the numerical model; we treat the decoherence as arising only from intragroup Ising interactions, which is correct at short times when the NV centers are spin-polarized. %At late times, the XY-8 data begins to curve downwards (grey shaded area), as the measurement starts to access the late-time ``random-walk'' regime. This behavior is caused by spin-flips in the bath of NVs, which are not captured by the pure Ising interactions in the numerical model; we thus do not include this portion of the measurement in the density estimation. 

With the above prescription, we compute a set of Ramsey signals~[Fig.~\ref{fig:densityEstimation}(a), dashed lines] as a function of areal density $n_\text{3D}^\text{NV} \cdot w$, which we compare against the XY-8 data~[Fig.~\ref{fig:densityEstimation}(a), orange points]. %The numerical curves plotted in Fig.~\ref{fig:densityEstimation}(a) are independent of the layer thickness $w$, for $w$ smaller than the upper bound of $w=8~\mathrm{nm}$ obtained from the SIMS measurement [Fig.~1(d) in main text]. 
The estimated areal density is thus $n_\text{3D}^\text{NV} \cdot w=19\pm2~\mathrm{ppm \cdot nm}$, %\sim 4.2(3)\times 10^{-3}~\mathrm{nm^{-2}}$, 
corresponding to a density $n_\text{3D}^\text{NV} = 3.2\pm0.3~\mathrm{ppm}$ assuming a $w = 6~\mathrm{nm}$-thick layer.

\subsubsection{P1 density}
We estimate the P1 density with a similar procedure, using DEER data instead of XY-8 data. We first remove the contribution due to NV-NV interactions from the DEER signal by subtracting an interpolation of the XY-8 data~[Fig.~\ref{fig:densityEstimation}(a)] from the raw DEER data. Then, we compare the measured early-time dynamics with numerically-computed curves for a range of P1 densities $n_\text{3D}^\text{P1}/3$, as shown in Figure~\ref{fig:densityEstimation}(b). Here, we include a factor of $1/3$ in the P1 density because the microwave tone $\omega_\text{P1}$ addresses only one-third of the P1 spins (the ``P1-1/3 group'') in our DEER measurement, which are separated by $\sim 100~$MHz from the four other groups due to the hyperfine interaction, \citeMethods{zu2021, hall:2016}.
By comparing the data and theory curves, we estimate an areal density of $n_\text{3D}^\text{P1} \cdot w=85\pm10~\mathrm{ppm \cdot nm} \sim 1.4(1)\times 10^{-2}~\mathrm{nm^{-2}}$. %corresponding to a density of $n_\text{3D}^\text{P1} = 14\pm2~\mathrm{ppm}$ assuming a $w = 6~\mathrm{nm}$-thick layer. %We again ignore the late-time $t\gg\tau_c$ region (gray shaded area) when comparing the measurement to the numerical curves. 

At fixed areal density, the numerics indicate that the DEER decoherence profile depends on the layer thickness, as shown in Figure~\ref{fig:densityEstimation}(d); the same dependence is not present in the XY-8 dynamics due to the relatively small density of NV centers~[Fig.~\ref{fig:densityEstimation}(c)]. Although this method does not yield nanometer resolution, our observations are inconsistent with a layer with $w>6$~nm, placing a more stringent bound on the thickness of the layer. The areal density $n_\text{3D}^\text{P1} \cdot w=85\pm10~\mathrm{ppm \cdot nm}$ corresponds to $n_\text{3D}^\text{P1} = 14\pm2~\mathrm{ppm}$, assuming a $w = 6~\mathrm{nm}$-thick layer.

Other spin-1/2 paramagnetic defects in diamond~\citeMethods{grinolds2014subnanometre} may have the same resonant frequency as the P1-1/3 group, causing a possible systematic error in our method for estimating the P1 density. To determine if such defects are present in sample S1, we measured the P1 spectrum and compared the relative integrated areas under the peaks for the P1-1/3, 1/4, and 1/12 groups, thus obtaining an estimate of the relative densities between P1 groups. As shown in Figure~\ref{fig:otherDefect}, the results agree with the expected ratios 1:0.75:0.25 and are consistent with a negligible contribution of non-P1 defects to the DEER signal.

\subsection{Sample S2}
A detailed characterization of the three-dimensional sample S2 is given in Ref.~\cite{eichhorn2019optimizing} (sample C041). Here, we describe the key properties relevant for the present study. The sample was grown by depositing a 32~nm diamond buffer layer, followed by a 500~nm nitrogen-doped layer (99$\% \ ^{15}$N), and finished with a 50~nm undoped diamond capping layer. Vacancies were created by irradiating with 145 keV electrons at a dosage of $10^{21}$~cm$^{-2}$, and vacancy diffusion was activated by annealing at 850$^\circ$C for 48 hours in an Ar/Cl atmosphere. The resulting NV density is $\sim0.4$~ppm, obtained through instantaneous diffusion measurements~\cite{eichhorn2019optimizing}. The P1 density is measured to be $\sim20$~ppm through a modified DEER sequence~\cite{eichhorn2019optimizing}. The average spacing between P1 centers ($\sim 4$ nm) is much smaller than the thickness of the nitrogen doped layer, ensuring three-dimensional behavior of the spin ensemble. 

\subsection{Samples S3 and S4}
Samples S3 and S4 used in this work are synthetic type-Ib single crystal diamonds (Element Six) with intrinsic substitutional $^{14}$N concentration $\sim100$~ppm (calibrated with an NV linewidth measurement \citeMethods{zu2021}). 
To create NV centers, the samples were first irradiated with electrons ($2$~MeV energy and $1\times10^{18}$~cm$^{-2}$ dosage) to generate vacancies. 
After irradiation, the diamonds were annealed in vacuum ($\sim10^{-6}$~Torr) with temperature $>800^\circ$C. The NV densities for both samples were measured to be  $\sim0.5$~ppm using a spin-locking measurement \citeMethods{zu2021}.

\section{Experimental Methods}

% This section is admittedly very repetitive. We could probably combine everything into one much shorter section. However, for now I keep it like this for maximum clarity.

\subsection{Experimental details for sample S1}
The delta-doped sample S1 was mounted in a scanning confocal microscope. For optical pumping and readout of the NV centers, about 100~$\mu$W of 532~nm light was directed through an oil-immersion objective (Nikon Plan Fluor 100x, NA 1.49). The NV fluorescence was separated from the green 532~nm light with a dichroic filter and collected on a fiber-coupled single-photon counter. A magnetic field $B$ was produced by a combination of three orthogonal electromagnetic coils and a permanent magnet, and aligned along one of the diamond crystal axes.% and set to $B = 270~$G, as calibrated via the spectrum of the NV centers. 
The microwaves used to drive magnetic dipole transitions for both NV and P1 centers were delivered via an Omega-shaped stripline with typical Rabi frequencies $\sim 2\pi \times 10~$MHz.

\subsubsection{DROID and Hartmann-Hahn sequences}
Here, we describe the pulse sequences used to perform the measurements in Figure~4 of the main text. In Figure 4(a), we compare the coherence times across different dynamical decoupling sequences, demonstrating that the longest coherence times are achieved when we decouple both on-site disorder and dipolar NV-NV dynamics using a DROID (Disorder-RObust Interaction Decoupling) sequence proposed by Choi et al.~\citeMethods{choi2020robust}.
%decouple and transform the native dipole-dipole interaction between NV centers to an effective Heisenberg interaction. 
% Besides the well-known DROID-60 sequence~\cite{choi2020robust, ZhouQuantumMetrologyStrongly2020}, we experimented with a few (out of many possible) variations on DROID-60 to achieve the best decoupling. 
To achieve the best decoupling, we experimented with a few variations on the well-known DROID-60 sequence~\cite{ZhouQuantumMetrologyStrongly2020}; these measurements are plotted in Figure~\ref{fig:FullDD}. The DROID-60 data exhibit a pronounced coherent oscillation (purple points), which we attribute mainly to errors in composite pulses formed by sequential $\pi/2$ rotations along different axes. The data exhibiting the longest coherence time (red points) are obtained using so-called Sequence H, shown in Figure~9 of Ref.~\citeMethods{choi2020robust}. We hypothesize that Sequence H behaves more predictably precisely because it eliminates composite pulses.  

In Figure~4(b - c), we demonstrate polarization transfer between NV and P1 spins using a Hartmann-Hahn sequence. Following an initial $\pi/2$ pulse, the NV centers are  spin-locked with Rabi frequency $\Omega_\text{NV}$, while the P1 centers are simultaneously driven with Rabi frequency $\Omega_\text{P1}$, for a duration $t_s$. A final $\pi/2$  pulse is applied before detection. When the two Rabi frequencies are equal $\Omega_\text{NV} = \Omega_\text{P1}$, the spins are resonant in the rotating frame, and spin-exchange interactions enhance the depolarization rate. The resonant depolarization data [Fig.~4(b) in main text, green curve] are well-fitted by the functional form
\begin{equation}
C(t) = e^{-t/T_{1\rho}}e^{-(t/\tau)^{D/2\alpha}},
\end{equation}
where the first factor captures phonon-limited exponential decay and the second factor captures the independent depolarization channel driven by spin-exchange interactions, with stretch power $\beta = 1/3$~\cite{choi2017depolarization}. We determine $T_{1\rho} = 1.05(3)~$ms from the NV spin-locking measurement [Fig.~4(b) in main text, orange curve]. 
%In the experiment, we first fixed the NV Rabi frequency at $(2\pi)\times5~\mathrm{MHz}$ and the spin-locking duration $\tau = 200~\mathrm{\mu s}$ and swept the Rabi frequency $\Omega_\text{P1}$. We observed that the NV polarization at $200~\mathrm{\mu s}$ is reduced when the Rabi frequencies of NV and P1 coincide [Fig]. Then, we fixed the Rabi frequencies such that NV and P1 centers are resonant, and measured the NV depolarization profile. Discuss stretch power.

\subsection{Experimental details for sample S2}
Sample S2 was mounted in a confocal microscope. For optical initialization and readout, about 350~$\mu$W of 532~nm light was directed through an air objective (Olympus UPLSA 40x, NA 0.95). The NV fluorescence was similarly separated from the 532~nm light with a dichroic mirror and directed onto a fiber-coupled avalanche photodiode. A permanent magnet produced a field of about 320~G at the location of the sample. The field was aligned along one of the NV axes, and alignment was demonstrated by maximizing the $^{15}$N nuclear polarization~\citeMethods{jacques2009dynamic}. Microwaves were delivered with a free-space rf antenna positioned over the sample.

\subsection{Experimental details for samples S3 and S4}
Samples S3 and S4 were mounted in a confocal microscope. For optical initialization and readout, about 3~mW of 532~nm light was directed through an air objective (Olympus LUCPLFLN, NA 0.6). The NV fluorescence was separated from the 532~nm light with a dichroic mirror and directed onto a fiber-coupled photodiode (Thorlabs). The magnetic field was produced with a electromagnet with field strength $\sim 174$~G ($\sim 275$~G) for sample S3 (S4). The field was aligned along one of the NV axes, and %alignment was demonstrated by making the resonances from the other 3 NV axes degenerate. 
microwaves were delivered using an Omega-shaped stripline with typical Rabi frequencies $\sim 2\pi\times 10~$MHz.

% Calibrating P1 spectrum/pulses?
% \subsubsection{NV Linewidth}
% The NV linewidth is measured with a pulsed ESR sequence. 
% \subsubsection{P1 Linewidth}
% The P1 linewidth is measured with a modified DEER sequence.

\subsection{Polychromatic drive}\label{sec:polychromaticDrive} 
The polychromatic drive was generated by phase-modulating the resonant P1 microwave tone~\citeMethods{joos2021protecting}. A random array of phase jumps $\Delta \theta$ was pre-generated and loaded onto an arbitrary waveform generator (AWG) controlling the IQ modulation ports of a signal generator. The linewidth of the drive $\delta\omega$ was controlled by fixing the standard deviation of the phase jumps $\sigma = \sqrt{\delta \omega \delta t}$ in the pre-generated array, where $1/\delta t = 1~$GS/s was the sampling rate of the AWG. %and confirmed experimentally via a Fourier transform of the measured SRS output. 
The power in the drive was calibrated by measuring Rabi oscillations of the P1 centers without modulating the phase, i.e. by setting $\delta\omega = 0$. %At fixed SRS output power and IQ modulation amplitude, we assume the integrated power remains fixed if we subsequently increase the linewidth.% to $\delta \omega$.

\section{Spin echo for samples S1 and S2 without polychromatic driving}\label{sec:undriven}

In \textit{Characterizing microscopic spin-flip dynamics}, we discussed spin echo measurements limited by NV-P1 interactions (as one would naively expect), and which exhibit an early-time stretch power of $\beta = 3D/2\alpha = 3/2$. These measurements were performed on samples S3 and S4 which exhibit a P1-to-NV density ratio of $\sim 200$. By contrast, spin echo measurements in samples S1 and S2, with P1-to-NV density ratios of $\sim 10$ and $\sim 40$, respectively, exhibit an early-time stretch $\beta = D/\alpha$~[Fig.~\ref{fig:undriven}], consistent with the prediction for a Ramsey measurement~[Table~1 in main text]. Here, we are discussing a ``canonical'' spin echo measurement with no polychromatic drive~[inset, Fig.~2(b) in main text], and thus this data is not in contradiction with that presented in Figure~3 of the main text.

A possible explanation for the observed early-time stretch $\beta = D/\alpha$ is that the spin echo signal is limited by NV-NV interactions, rather than by NV-P1 interactions. In order to understand this limitation, it is important to realize that the measured spin echo signal actually contains at least two contributions: (i) the expected spin echo signal from NV-P1 interactions, arising because the intermediate $\pi$-pulse decouples the NVs from any quasi-static P1 contribution; (ii) a Ramsey signal from NV interactions with other NVs, arising because these NVs are flipped together by the $\pi$-pulse, and the intra-group Ising interactions are not decoupled. %it is not possible to echo away interactions between NVs with the same resonance frequency (

%Two pieces of evidence support our hypothesis that NV-NV interactions limit the spin-echo coherence in samples S1 and S2. %First, for sample S1, fitting the XY-8~(directly characterizing the NV-NV interaction) [orange data, Fig.~\ref{fig:densityEstimation}(a)] and spin echo signals [Fig.~\ref{fig:undriven}(a)] using $\chi^\text{DEER}$ yields the same fitted NV-NV correlation time $\tau_c^\text{NV} = 50~\mu$s, as well as similar (within a factor of two) 1/e decay timescales. %, where $\tau_c^\text{NV}$ is the correlation time of the NV centers in the Ramsey measurement. 
%In fact, the two data sets are extremely similar, up to a constant factor $\sim 2$ in the decay timescale, which could be attributed to differences between the two pulse sequences (XY-8 versus spin echo). 
Our hypothesis that NV-NV interactions limit the spin-echo coherence in sample S1 is supported by the fact that a stretch power of $\beta = 3D/2\alpha = 1$ can in fact be observed in spin echo data if the environment is made significantly noisier, e.g. by reversibly worsening the quality of the diamond surface~[Fig.~\ref{fig:undriven}(b), (green points)]. A subsequent three-acid clean restores the original $\beta = 2/3$ stretch power~[Fig.~\ref{fig:undriven}(a), (red points)]. 
%Finally, we have observed experimentally that two samples (S3, S4) with NV to P1 density ratios $\sim 200$ exhibit $\beta = 3D/2\alpha$, whereas three samples (S1, S2) with lower P1 to NV density ratios $\sim 10, 40$ respectively exhibit $\beta = D/\alpha$. While we have not studied the stretch power systematically as a function of the P1-to-NV density ratio, our four existing results agree qualitatively with our hypothesis. 

\section{Data Analysis and Fitting}
\subsection{Normalization of decoherence data}\label{sec:Norm}

The coherence of the NV spins is read out via the population imbalance between $\{ \ket{0}, \ket{-1}\}$ states. The maximum measured contrast $\lesssim 8~\%$ is proportional --- {\it not} equal --- to the normalized coherence $C(t)$. To see a physically-meaningful stretch power in our log-log plots of the data [Fig.~2, Fig.~3 in main text], it is necessary to normalize the data by an appropriate value that captures our best approximation of the $t=0$ time point for the DEER and spin echo measurements.  

\subsubsection{Samples S1, S3, and S4 $t=0$ measurement}
For a given pulse sequence (e.g. Ramsey or spin echo) and fixed measurement duration $t$, we perform a differential readout of the populations in the $\ket{0}$ and $\ket{-1}$ spin states of the NV, which mitigates the effect of NV and P1 charge dynamics induced by the laser initialization and readout pulses. As depicted schematically in Fig.~\ref{fig:sequence}, we allow the NV charge dynamics to reach steady state (I) before applying an optical pumping pulse (II). Subsequently, we apply microwave pulses to both the NV and P1 spins (e.g. Ramsey or spin echo pulse sequences shown in Figs.~2-3 of the main text) (III). Finally, we detect the NV fluorescence (IV) to measure the NV population in $\ket{0}$, obtaining a signal $S_0$. We repeat the same sequence a second time, with one additional $\pi$-pulse before detection to measure the NV population in $\ket{-1}$, obtaining a signal $S_{-1}$. The raw contrast, $C_\text{raw}$, at time $t$ is then computed as $C_\text{raw}(t) \equiv [S_0(t) - S_{-1}(t)]/S_0(t)$, and is typically $\lesssim 8\%$. %In principle, $R_0(t)$ should be independent of $t$ but can fluctuate due to e.g. optical power drifts. 
We normalize the raw contrast to the $t=0$ measurement to obtain the normalized coherence, $C(t)$, defined in the main text:
\begin{equation}
    C(t) = C_\text{raw}(t)/C_\text{raw}(t=0).
\end{equation}

\subsubsection{Sample S2 $t=0$ measurement}
For sample S2, we have an early-time, rather than a $t=0$, measurement at $t=320~$ns for spin echo and DEER sequences. Because the DEER signal decays on a much faster timescale than the spin echo signal, we normalize both datasets to the earliest-time spin echo measurement.

\subsection{Data analysis  for Figure~3 in main text}
We separate our discussion of the data analysis relevant to Fig.~3 of the main text into two parts: First, we discuss how comparing the $D=2$ and $D=3$ best-fits to the DEER measurements enable us to identify the dimensionality of the underlying spin system.
Second, armed with the fitted dimensionality, we fit spin echo and DEER measurements simultaneously to Eqn.~5 of the main text to extract the correlation time $\tau_c$ of the P1 system.
We note that, except for the $t=0$ normalization point (Sec.~\ref{sec:Norm}), we only consider data at times  $t>0.5~\mathrm{\mu s}$ to mitigate any effects of early-time coherent oscillations caused by the hyperfine coupling between the NV and its host nitrogen nuclear spin.
% We highlight these features within the context of the driven data where the Gaussian-Markov nature of the bath is stipulated by the applied incoherent drive.

\subsubsection{Determining the dimensionality of the system}

In order to determine the dimensionality of the different samples S1 and S2, we focus on the DEER signal, where the stretch power is given by $\beta = D/\alpha$ in the early-time ballistic regime and $\beta = D/2\alpha$ in the late-time random walk regime.
Employing both Gaussian and Markovian assumptions, a closed form for the decoherence can be obtained as~\cite{SM}:
\begin{equation}
    \label{eq:DEER}
    C^\text{DEER}(t) = e^{-A[\chi^\text{DEER}(t)]^{D/2\alpha}},\\
\end{equation}
where $\chi^\text{DEER}$ is defined in Eqn.~5 of the main text.

Armed with Eqn.~\ref{eq:DEER}, we consider the decoherence dynamics for different powers of the polychromatic drive for both $D=2$ and $D=3$ (with $\alpha = 3$, as per the dipolar interaction).
We compare the reduced $\chi_{\mathrm{fit}}^2$ goodness-of-fit parameters for the two values of $D$, and demonstrate that stretch power analysis of the main text indeed agrees with the dimensionality that best explains the observed DEER data. Changing the dimension $D$ does not change the  number of degrees of freedom in the fit, so a direct comparison of $\chi_{\mathrm{fit}}^2$ is meaningful. Our results are summarized in Fig.~\ref{fig:dim}, where we observe that for sample S1 indeed the $D=2$ fitting leads to a smaller  $\chi_{\mathrm{fit}}^2$, while for sample S2 the data is best captured by $D=3$ [Fig.~\ref{fig:dim}]. Independently fitting both the extracted signal $C(t)$ as well to its negative logarithm $-\log C(t)$ yields the same conclusions. This analysis complements the discussion in the main text in terms of the early-time and late-time stretch power of the decay.
%
%A few remarks are in order. First, note that changing the dimension $D$ does not change the  number of degrees of freedom in the fit, so a direct comparison of $\chi_{\mathrm{fit}}^2$ is meaningful. 

\subsubsection{Extracting the correlation time $\tau_c$}
Having determined the dimensionality of samples S1 and S2, we now turn to characterizing the correlation times of the P1 spin systems in these samples.
%Unfortunately, in this case, the DEER signal alone does not yield a robust measurement of the correlation time; %the early and late-time stretch powers are similar and 
%the crossover region is too broad.
To robustly extract $\tau_c$, we perform a simultaneous fit to both the DEER signal with Eqn.~\ref{eq:DEER} and the spin echo signal with 
% \begin{sube quations}
\begin{equation}
\label{eq:ECHO}
C^\text{SE}(t) = e^{-A[\chi^\text{SE}(t)]^{D/2\alpha}},
\end{equation}
% \end{subequations}
assuming a single amplitude $A$ and correlation time $\tau_c$ for both normalized signals. Here, $\chi^{\mathrm{DEER/SE}}$ depends on $\tau_c$ as defined in Eqn.~5 of the main text. 

In order to carefully evaluate the uncertainty in the extracted correlation time, we take particular care to propagate the uncertainty in the $t=0$ data used to normalize the raw contrast, i.e. $C_\text{raw}(t=0)$~[Sec.~\ref{sec:Norm}].
Owing to the two  normalization methods for samples S1 and S2 [Sec.~\ref{sec:Norm}], we estimate the uncertainty in two different ways:
\begin{itemize}
    \item For samples S1, S3, and S4, we consider fluctuations of the normalization value, $C_\text{raw}(t=0)$, by $\pm 10\%$. This is meant to account for a possible effect of the hyperfine interaction in this data point, as well as any additional systematic error.
    \item For sample S2, we first compute a linear interpolation of the early time spin echo decoherence to $t=0$. 
    We then sample the normalization uniformly between this extrapolated value and the earliest spin echo value.
\end{itemize}

By sampling over the possible values of $C_\text{raw}(t=0)$, we build a distribution over the extracted values of $\tau_c$ fitting to both the coherence, $C(t)$, and its logarithm, $-\log~C(t)$.
The reported values in Fig.~3(d, e) correspond to the mean and standard deviation evaluated over this distribution.

We end this section by commenting that, as the drive strength is reduced the spin echo signal looks increasingly similar to the undriven spin echo data [Fig.~\ref{fig:undriven}], i.e. the early time stretch changes from $\beta = 3D/2\alpha$ to $\beta = D/\alpha$; our explanation for this observed stretch is given in Sec.~\ref{sec:undriven}. 
The deviation from the expected functional form for the decoherence leads to a large uncertainty in the extracted correlation time. The data also deviates from the model for larger drive strengths, e.g. $\Omega = 2\pi\times 4.05~$MHz, $\delta \omega = 2\pi\times 20~$MHz, where our assumption that $\delta \omega \gg \Omega$ is no longer valid [Fig.~\ref{fig:fastDriven}].

\section{Extended Data}
\setcounter{figure}{0}
\renewcommand{\thefigure}{E\arabic{figure}}
\renewcommand{\thetable}{E\arabic{table}}
\begin{figure}[H]
    \centering
    \includegraphics[width = 5.8in]{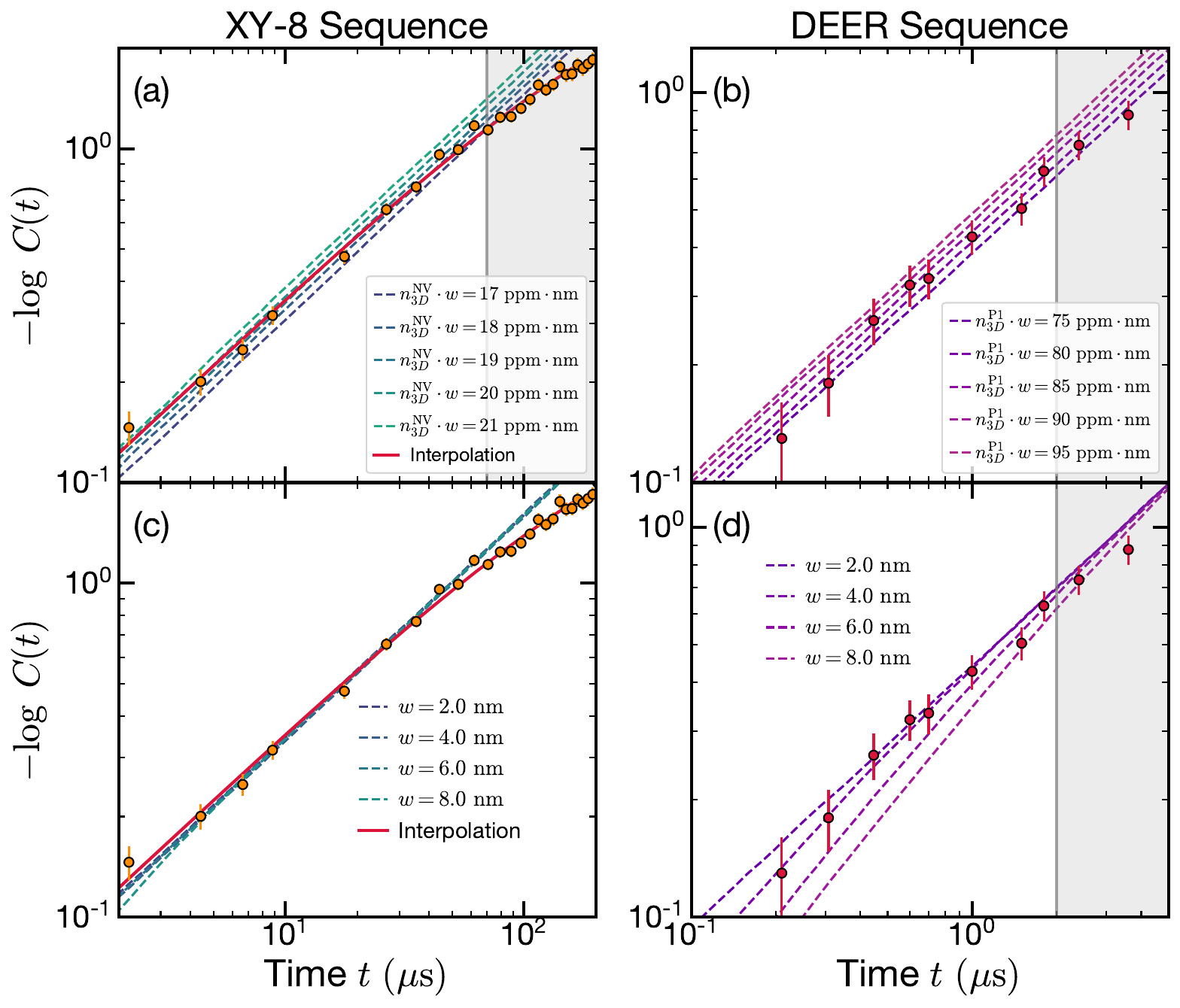}
    \caption{
    (a-b) NV and P1 areal densities. Computed dynamics (dashed lines) for early-time Ramsey decoherence caused by NV-NV interactions (a) and NV-P1 interactions (b), as a function of the NV areal density $n_\text{3D}^\text{NV} \cdot w$ and the P1 areal density  $n_\text{3D}^\text{P1} \cdot w$, respectively.
    We compare these numerical results against the measured decoherence dynamics obtained via the XY-8 sequence (orange points, a) and the DEER sequence [after removing the NV contribution via the red interpolation in (a)] (purple points, b) to obtain the areal density of defects in sample S1. We estimate the areal density of NV centers to be $n_\text{3D}^\text{NV} \cdot w=19\pm2~\mathrm{ppm \cdot nm}$ and the P1 density to be $n_\text{3D}^\text{P1} \cdot w=85\pm10~\mathrm{ppm \cdot nm}$.
    At late times (grey shaded regions), the noise dynamics approach an incoherent random walk and should not be used to compute the density within this analysis, because the numerics do not include flip-flop interactions.
    (c-d) Effect of finite-thickness layer. At fixed areal density $19~\mathrm{ppm\cdot nm}$, choosing different layer widths $w$ does not affect the computed dynamics (dashed curves, c). For higher density P1 centers $n_\text{3D}^\text{P1} = 85~\mathrm{ppm\cdot nm}$, the finite thickness of the layer can induce a sizable effect on the DEER decoherence dynamics (purple points, d) at early times. Numerical calculations are plotted as dotted lines. %performed with areal density of $24$ and $85~\mathrm{ppm\cdot nm}$ for panels (c) and (d), respectively.
    }
    \label{fig:densityEstimation}
\end{figure}

\begin{figure}[H]
    \centering
    \includegraphics[width = 3in]{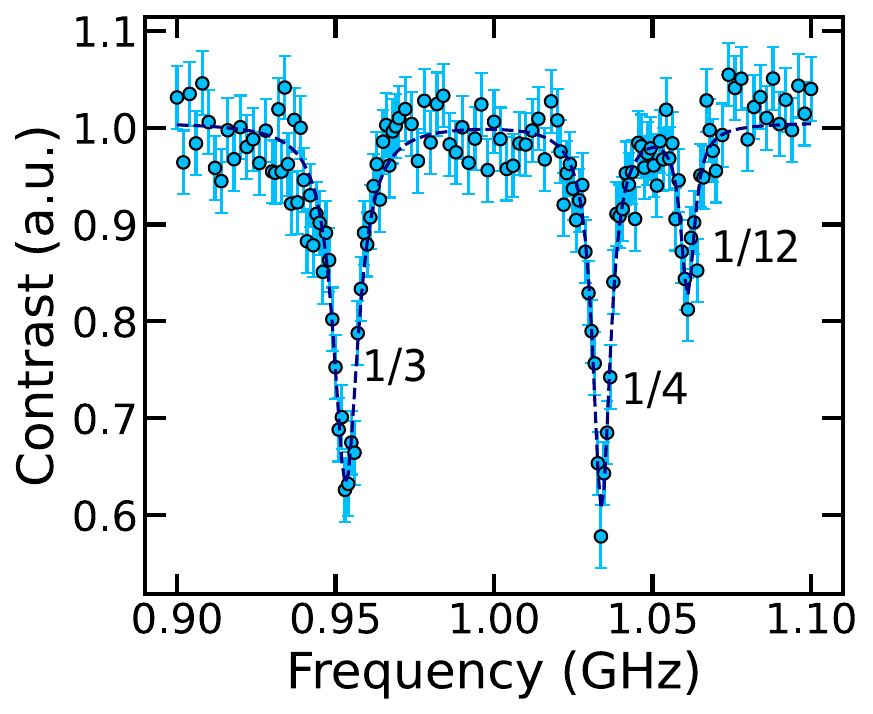}
    \caption{Relative density extraction for three of the five different P1 groups. The P1 spectrum is fit to a sum of three Lorentzian curves (dashed line). The relative areas of the three dips are 1 : 0.74 : 0.26, which is consistent with the expected ratio 1 : 0.75 : 0.25. %Hence, we observe no convincing evidence of the existence of other $g=2$ defects in sample S1. 
    }
    \label{fig:otherDefect}
\end{figure}

\begin{table*}[H]
    \centering
    \begin{tabular}{c||c|c|c|c}  \hline \hline 
         \makecell{Parameter} & \makecell{S1} &  \makecell{S2} & \makecell{S3} &\makecell{S4} \\ \hline \hline 
         P1 Density             & 85(10)~ppm $\cdot$ nm & 20(1)~ppm & $\sim$100~ppm & $\sim$100~ppm\\ \hline
         NV Density             & 24(2)~ppm $\cdot$ nm & 0.43(1)~ppm & $\sim$0.5~ppm & $\sim$0.5~ppm\\ \hline
         Diamond cut            & [100] & [100] & [111] & [100]\\ \hline
         Nitrogen isotope       & 14 & 15 & 14 & 14\\ \hline
         Isotopically purified?       & Yes & Yes & No & No\\ \hline
         Additional Comments      & CVD grown & CVD grown, Ref.~\cite{eichhorn2019optimizing} & Type Ib, Ref.~\citeMethods{zu2021} & Type Ib\\ \hline \hline
    \end{tabular}
    \caption{Summary of sample parameters.}
    \label{tab:table1}
\end{table*}

\begin{figure}[H]
    \centering
    \includegraphics[width = 4in]{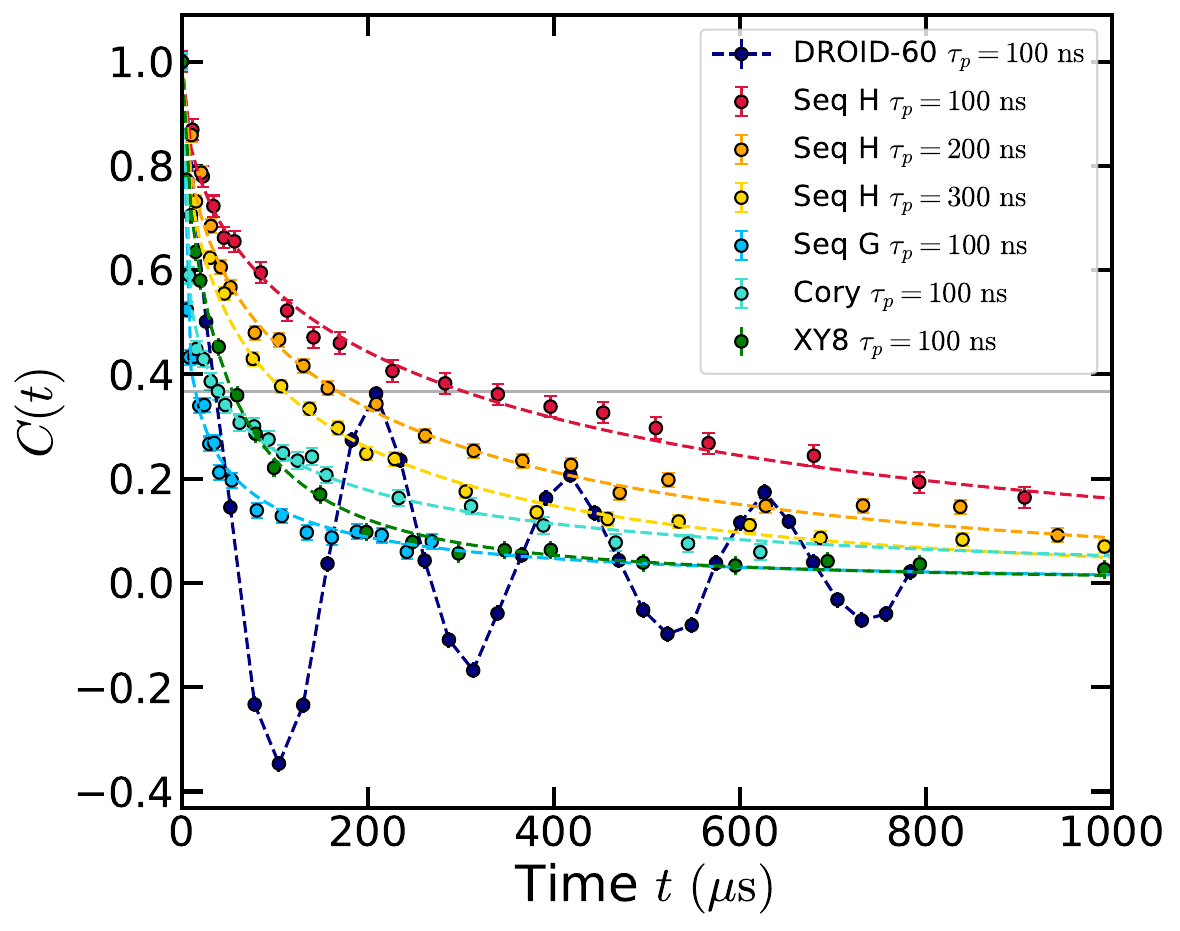}
    \caption{The measured decoherence profiles of variations on DROID-60~\protect\citeMethods{choi2020robust}. Due to imperfections in our composite microwave pulses, pulse error accumulates coherently in the DROID-60 sequence and a pronounced oscillation is observed (purple points). To avoid such oscillations, we implement sequences that do not require composite pulses, see e.g. Seqs. A, H, G in Fig. 9 of Ref.~\protect\citeMethods{choi2020robust} The data exhibiting the longest coherence time (Seq. H, $\tau_p = 100~$ns, red points) are also shown in Figure~4(a) of the main text.}
    \label{fig:FullDD}
\end{figure}

\begin{figure}[H]
    \centering
    \includegraphics[width = \textwidth]{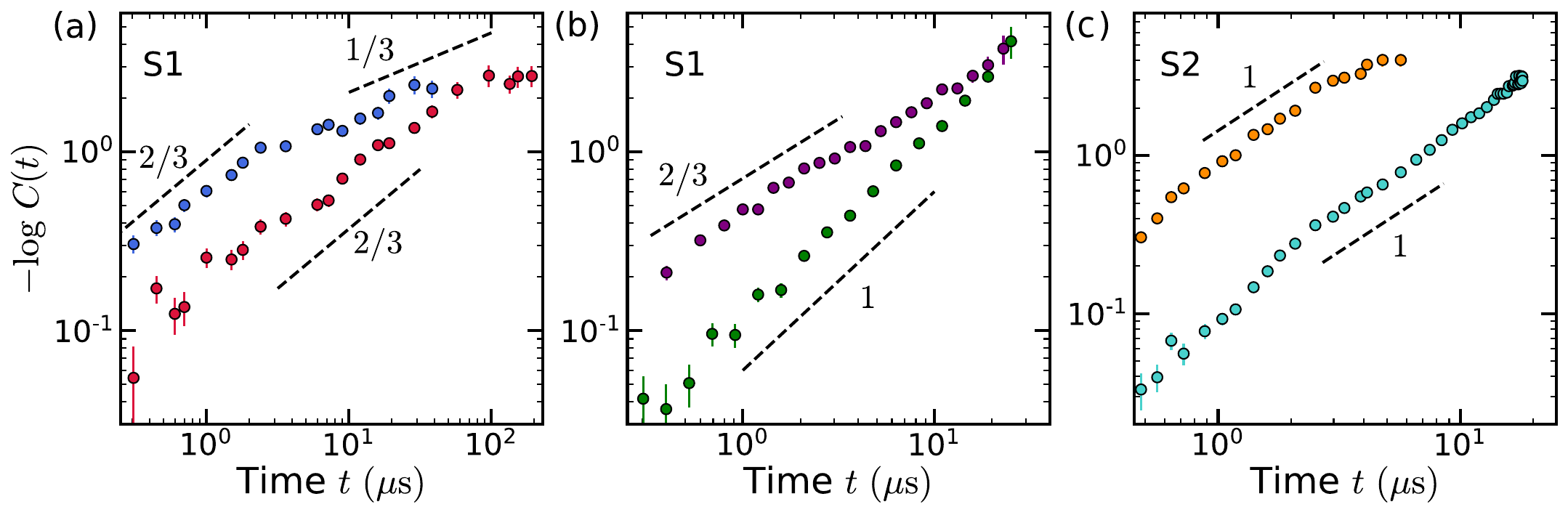}
    \caption{Undriven DEER and spin echo data. %for sample S1 (a, clean surface; b, modified surface) and sample S2 (c).
    (a) In sample S1, with a clean surface, both the DEER (blue) and spin echo  (red) data exhibit a stretch power $\beta = 2/3$. (b) After worsening the surface quality, the spin bath becomes noisier and we observe the expected $\beta = 1$ stretch power in the echo data (green); in the DEER data (purple) the correlation time $\tau_c$ increases but the stretch power is unchanged. (c) As in panel (a), the spin echo data (teal) for sample S2, presumably limited by NV-NV interactions rather than the bath, exhibit the same stretch power $\beta = 1$ as the DEER data. The DEER data in (a, c) are also plotted in Fig.~2(a) of the main text.
    }
    \label{fig:undriven}
\end{figure}

\begin{figure}[H]
    \centering
    \includegraphics[width=4in]{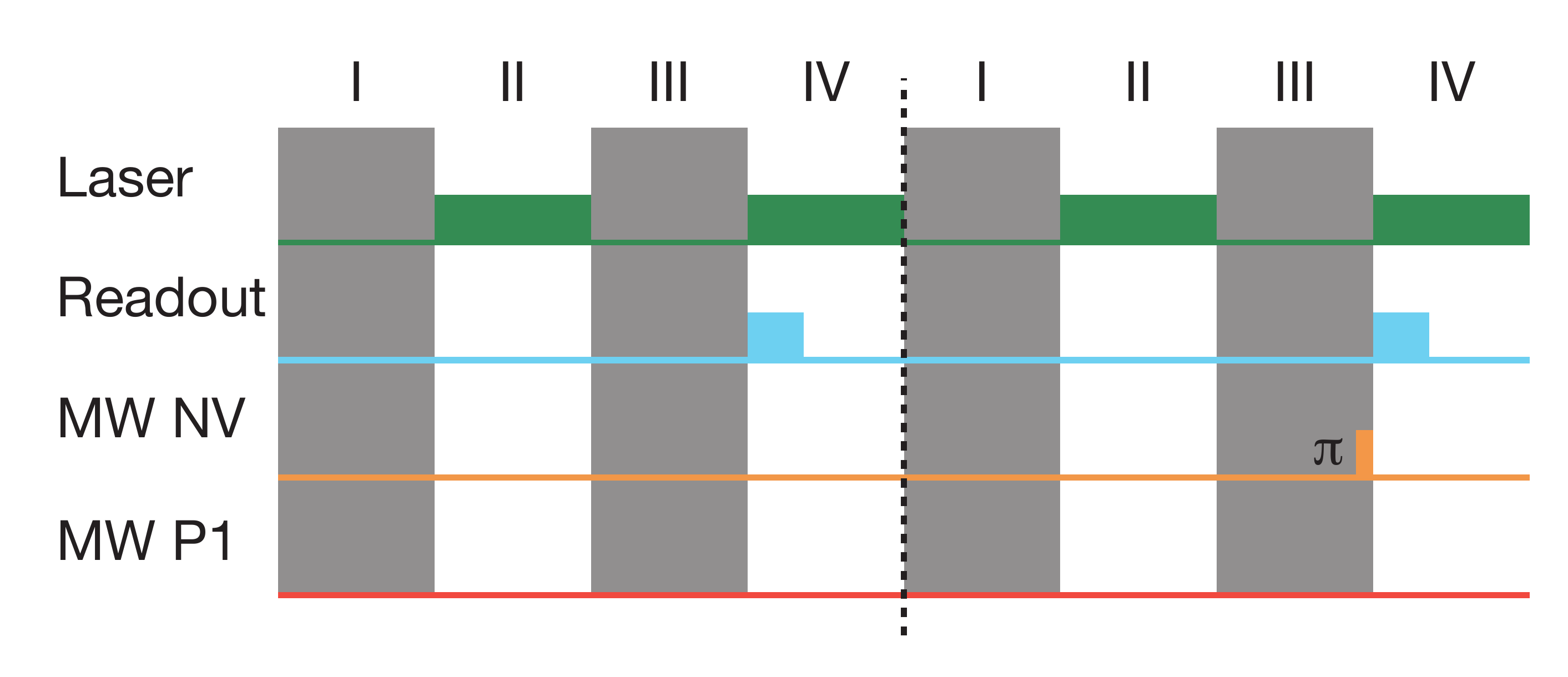}
    \caption{Experiment sequence schematic for differential measurement.}
    \label{fig:sequence}
\end{figure}

\begin{figure}[H]
    \centering
    \includegraphics[width=5.7in]{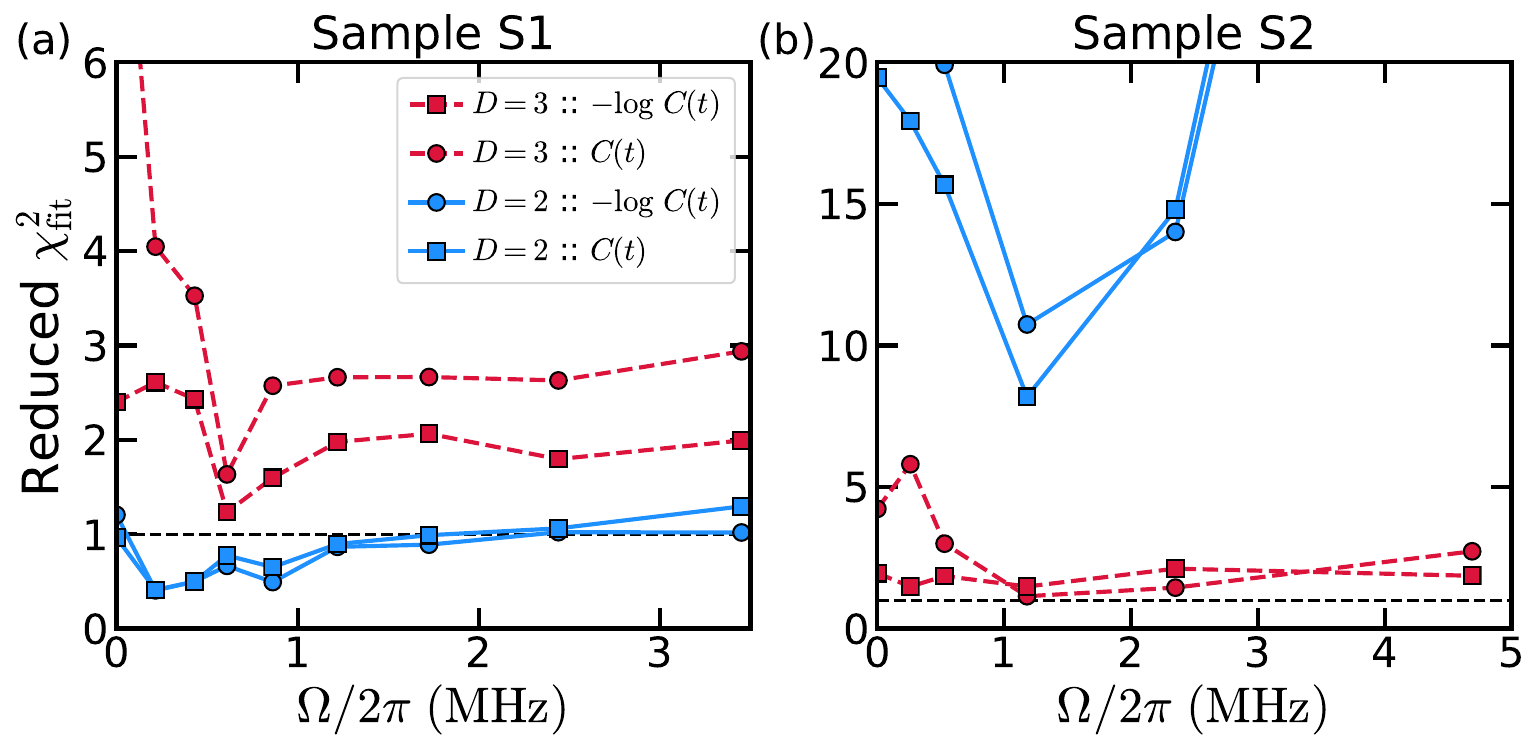}
    \caption{
    Reduced $\chi_\mathrm{fit}^2$ for fits to the DEER measurements on samples S1 (a) and S2 (b) as a function of the incoherent driving strength, for four different fit models: we fit to both dimensions ($D=2$ and $D=3$) using the decoherence directly [$C(t)$] or its negative logarithm [$-\log~C(t)$].
    For samples S1 and S2, fits in both log and linear space show that $D=2$ and $D=3$ respectively better capture the data. The consistency of this result across a range of drive powers highlights our ability to distinguish the dimensionality of the sample directly from the decoherence dynamics. 
    }
    \label{fig:dim}
\end{figure}

\begin{figure}[H]
    \centering
    \includegraphics[width = 6in]{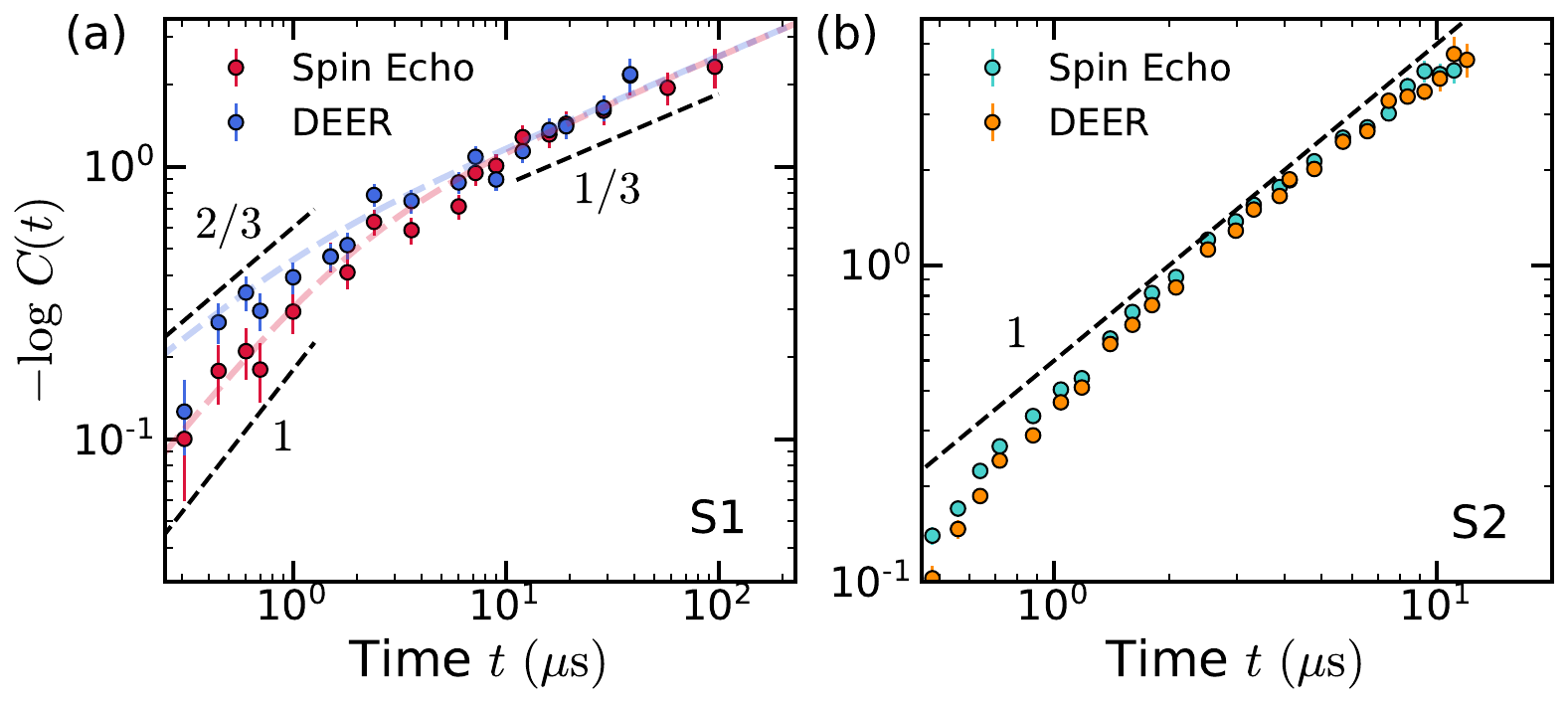}
    \caption{DEER and spin echo for sample S1 (a) and sample S2 (b), under a fast incoherent drive ($\Omega = 2\pi \times 3.45~\mathrm{MHz}$ and $\Omega = 2\pi \times 4.05~\mathrm{MHz}$, respectively). The data obtained for sample S2 plotted in (b) does not exhibit the correct ``random-walk'' regime stretch power of 1/2, even though the DEER and spin echo signals overlap at all measured times. %We believe this is caused by a large drive strength $\Omega$ relative to the linewidth $\delta \omega$.
    }
    \label{fig:fastDriven}
\end{figure}

\newpage
\section*{Data Availability}
Data supporting the findings of this paper are available from the corresponding authors upon request. Source data are provided with this paper.
\section*{Code Availability (if relevant)}
Code developed for the data analysis and visualization is available from the corresponding author upon request.
% \section*{Methods-only reference}
\bibliographystyleMethods{naturemag}
\bibliographyMethods{methods_finalSubmission}

\newpage

\end{document}

\title{Supplementary Information:\\
Probing many-body dynamics in a two dimensional dipolar spin ensemble}

\maketitle
In this supplement, we provide additional details on the results presented in the ``Theoretical framework for decoherence dynamics induced by many-body noise" section of the main text.
In Sec.~\ref{sec:Gauss-Markov}, we derive the decoherence dynamics within a semi-classical description that assumes Gaussian fluctuations.
This presentation collects many ideas found in previous works \cite{dobrovitski:2009, yang:2017, anderson1953exchange,  klauder1962spectral, hu1974theory, salikhov1981theory, fel1996configurational, cucchietti2005decoherence, de_Sousa_2009, wang2012comparison, kubo2012statistical, lukasz2008how, wang2013spin, hu1974theory, glasbeek1981phase,salikhov1981theory, fel1996configurational}. We aim to provide a clear, self-contained discussion and to derive a highly general expression for the decoherence dynamics which can be applied to a range of spin systems. 
In Sec.~\ref{sec:quantum}, we provide a complementary quantum mechanical description of the same decoherence dynamics, which enables us to frame the distinction between Gauss-Markov and telegraph noise in terms of the delocalization of quantum information.

Throughout this supplement, we consider the same setup as in the main text; namely, a probe spin-1/2 $\hat{s}_p$ interacting via long-range, $1/r^{\alpha}$, Ising interactions with system spins $\hat{s}_i$:
\begin{equation}
H_z =\sum_i \frac{ J_z  g_i}{r_i^\alpha}\hat{s}^z_p \hat{s}^z_i,
\label{eq:Ising}
\end{equation}
where we have explicitly separated the overall interaction strength $J_z$ from any angular dependence $g_i \sim \mathcal{O}(1)$. 

We are interested in the decoherence dynamics where the probe spin is initially prepared along the $\hat{x}$-axis of the Bloch sphere and the Ising interactions [Eqn.~\ref{eq:Ising}] cause its precession in the equatorial plane.
The coherence of the spin is defined as the (normalized) projection of the spin into the $\hat{x}$-axis at a later time, $C(t) = 2\braket{\hat{s}^x_p(t)}$. 

During the evolution, a decoupling sequence can be applied which flips the probe spin.
In the rotating frame, the decoupling sequence can be thought of as changing the sign of the Ising interaction.
This can be straightforwardly accounted for by introducing a function, $\eta(t';t)$, which captures the sign of $H_z$ at time $t'$ for a pulse sequence of duration $t$. For example, for a Ramsey measurement $\eta_\text{R}(t';t) = 1$, while for spin echo
\begin{equation}
    \eta_{\text{SE}}(t';t) = 
    \begin{cases} 1 & 0<t'<t/2 \\
    -1 & t/2 < t' < t.
\end{cases}
\end{equation}

We note that this analysis focuses only on Ising interactions and thus ignores the effects of depolarization on the decoherence dynamics. 
This is a natural assumption for NV-P1 systems where the two spin defects are far detuned ($\sim \mathrm{GHz}$) and inter-species spin-exchange interactions ($\sim \mathrm{MHz}$) are highly off-resonant and suppressed~\cite{zu2021}.

\section{Derivation of the decoherence dynamics assuming Gauss-Markov noise}
\label{sec:Gauss-Markov}

In this section we analyze the decoherence problem by treating the system spins as fluctuating classical variables $\hat{s}^z_i(t) \to s^z_i(t)$.
As such, instead of analysing the details of the dynamics of $\hat{s}^z_i$, we can relate the decoherence of the probe spin to the statistical properties of the fluctuating classical variables $s^z_i(t)$.

Within this framework, every trajectory of $s^z_i(t')$ induces a precession angle $\phi(t)$. The observed coherence is then given by the ensemble average of the coherence over all trajectories:
\begin{equation}\label{eq:Sx1}
C = 2\langle \hat{s}^x_p(t) \rangle = \left \langle \mathrm{Re}\left[e^{-i\phi(t)}\right] \right \rangle, \quad\text{with} \quad 
\phi(t) = \sum_{j} \frac{J_zg_j}{r_{j}^\alpha} \int_0^t \eta(t';t)s^z_i(t') dt',
\end{equation}
where $\langle \ldots \rangle$ denotes the average over the ensemble of trajectories.

% \sout{Here $\phi(t)$ is the accumulated phase of the probe spin under the Ising interactions with the system for an arbitrary pulse sequence $\eta(t';t)$, and generalizes the phase $\phi(t)$ defined in the main text. 
% Eqn.~\ref{eq:Sx1} describes the decoherence dynamics following a single trajectory of the system spins $s^z_i(t)$. 
% In practice, the system's spins are in a fully mixed state, where the initial state $s^z_i(t=0)$ is completely random and thus the dynamics $s^z_i(t)$ can be treated as a random variable. Thus, $\phi(t)$ will also correspond to a random variable.}

In order to proceed from Eqn.~\ref{eq:Sx1} into a closed form solution for the decoherence dynamics, we must relate the system dynamics to the statistical properties of $s^z_i(t)$ and $\phi(t)$; for simplicity, and following previous literature \cite{dobrovitski:2009,yang:2017, herzog1956transient, hanson2008coherent, de2010universal, wang2012comparison}, we assume that both $\phi(t)$ and $s^z_i(t)$ are accurately captured by Gauss-Markov processes.
We postpone a detailed discussion about the validity of this assumption to a later section.
%
% To determine the statistical properties of $s^z_i(t)$ and $\phi(t)$ requires knowledge of the dynamics; for simplicity we assume that both represent Gauss-Markov processes. 
% {\color{red}While this is certainly a reasonable and natural assumption inspired by the central limit theorem, we postpone a detailed discussion about its subtlety until the next section.}
%\sout{In many cases, this is a reasonable assumption for $\phi(t)$ because of the central limit theorem.}
%
This simplification allows us to obtain the decoherence dynamics in terms of the variance of $\phi(t)$:
\begin{align}
C(t) = \langle \mathrm{Re}[e^{-i\phi(t)}] \rangle = e^{-\langle \phi^2 \rangle / 2} &= \exp\left\{-\frac{1}{2}\left\langle \left\vert\sum_{i}\frac{J_z g_i \int_0^t\eta(t';t)s^z_i(t') dt'}{r_{i}^\alpha}\right\vert ^2 \right\rangle \right\} \notag \\
&= \prod_{i}\exp\left\{-\frac{1}{2}\left[ \frac{J_z|g_i| \chi(t)^{\frac{1}{2}}}{2r_{i}^\alpha}\right]^2  \right\}.
\label{eq:Sx2}
\end{align}
Here, we assume that each spin is independent and define $\chi(t)$ as follows:
%as the single spin two-point correlation function
\begin{equation}
\chi(t) \equiv 4\left\langle\left[ \int_0^t\eta(t';t)s^z_i(t') dt'\right]^2 \right\rangle = \int_0^t dt' \int_0^t dt'' \eta(t';t)\eta(t'';t)\left\langle 4s^z_i(t')s^z_i(t'') \right\rangle.
\label{eq:chi_corr}
\end{equation}
The assumption of Gauss-Markov noise enables us to write down the two-point correlation function in terms of a decaying exponential with time scale $\tau_c$:
\begin{equation}
    \xi(\tau)\equiv \left\langle 4s^z_i(t)s^z_i(t+\tau) \right\rangle = \mathrm{e}^{-|\tau|/\tau_c}.
\end{equation}
% $\chi(t)$ can be evaluated analytically upon including the corresponding correlation time $\tau_c$:
% \begin{equation}
%     \xi(\tau)\equiv \left\langle 4s^z_i(t)s^z_i(t+\tau) \right\rangle = \mathrm{e}^{-|\tau|/\tau_c}.
% \end{equation}

The factor of $4$ ensures that the correlation function $\xi(\tau)$ is normalized for spin-1/2 particles when $\tau=0$. 
% where $\tau_c$ is the correlation time of the spins $s_i$, and we use the fact that $\left\langle 4s^z_i(t_0)s^z_i(t_0) \right\rangle = 1$ at infinite temperature. %We note that this exponential form of correlation is motivated by the decay profile of depolarization in single-NV experiment.
Depending on the specific pulse sequence applied to the system spins, captured by $\eta(t';t)$, we can analytically obtain the expression for $\chi(t)$~[Table~\ref{tab:chi}].

A few remarks are in order.
First, $\chi(t)$ has an intuitive and straightforward interpretation in Fourier space.
Defining $f(\omega;t)$ as the Fourier transform of $\eta(t';t)$ and $S(\omega)$ as the Fourier transform of $\xi(\tau)$, $\chi(t)$ can be rewritten as:
\begin{equation}
\label{eq:chi_corr_freq}
\chi(t) = \int d\omega~|f(\omega; t)|^2 S(\omega),
\end{equation}
which recovers Eqn.~2 of the main text.
In this language, the role of the pulse sequence becomes clear: given the noise spectrum $S(\omega)$ of the spin dynamics $s_i^z(t)$, the pulse sequence acts as a filter function $f(\omega;t)$.
Changing the pulse sequence modifies the probe spin's sensitivity to different frequency components of the noise.

Second, for a single realization, the time-dependence of the decoherence dynamics is entirely determined by $\chi(t)$.
If $\chi(t)$ only depends on $\tau_c$ and the pulse sequence, the decoherence dynamics is not sensitive to the dimensionality of the spin system. The ability to probe the dimensionality of the spin system arises from the interplay between the positional disorder and the power-law interactions, as discussed in detail in the next section.

% \begin{table}
%   \centering
%  \begin{tabular}{c|c|c|c|c} 
%     \hline 
%      Sequence & $\eta(t';t)$ & $\chi(t)$ & \makecell{short-time\\ ($t\ll\tau_c$)} &\makecell{long-time\\ ($t\gg\tau_c$)}\\ 
%     \hline
%     Ramsey& 1 & $2\tau_ct-2\tau_c^2 \left(1-\mathrm{e}^{-\frac{t}{\tau_c}}\right)$& $t^2-\frac{t^3}{3\tau_c}$ & $2\tau_ct-2\tau_c^2$ \\ 
%     \hline
%     Echo & \begin{cases}1 \qquad 0\leq t'<t/2\\
%     -1 \quad t/2\leq t'<t\end{cases}& $2\tau_ct-2\tau_c^2\left(3+\mathrm{e}^{-\frac{t}{\tau_c}}-4\mathrm{e}^{-\frac{t}{2\tau_c}}\right)$ & $\frac{t^3}{6\tau_c}$ & $2\tau_ct-6\tau_c^2$ \\
%     % \hline
%     % XY8-1 & $\frac{\tau_p^2 }{12\tau}t+\frac{\tau_p^2}{4}(1-\mathrm{e}^{-\frac{t}{\tau_c}})$ & $\frac{\tau_p^2 }{3\tau_c}t$ & $\frac{\tau_p^2 }{12\tau_c}t+\frac{\tau_p^2}{4}$ \\
%     \hline
%     XY-8 & \begin{cases}1 \qquad (m-\frac{1}{4})\tau_p\leq t'<(m+\frac{1}{4})\tau_p\\
%     -1 \quad (m+\frac{1}{4})\tau_p\leq t'<(m+\frac{3}{4})\tau_p\end{cases} & $\frac{\tau_p^2}{12\tau_c}t$ & $\frac{\tau_p^2}{12\tau_c}t$ & $\frac{\tau_p^2}{12\tau_c}t$ \\
%     \hline
%   \end{tabular}
%   \caption{Expressions of $\chi(t)$ for Ramsey/DEER, spin echo, and XY-8. In XY-8, we assume the inter-pulse spacing $\tau_p\ll\tau_c$. 
%     }
% \label{tab:chi}
% \end{table}
\renewcommand{\arraystretch}{2}
\setlength{\tabcolsep}{4pt}
\begin{table*}[t]
    \centering
    \begin{tabular}{c||c|c|c|c}  \hline \hline 
         \makecell{Sequence} & \makecell{$\eta(t';t)$} &  \makecell{$\chi(t)$} & \makecell{Short-time\\$t\ll\tau_c$} &\makecell{Long-time\\$t\gg\tau_c$} \\ \hline \hline 
         \makecell{Ramsey\\(DEER)}& 1 & $2\tau_ct-2\tau_c^2 \left(1-\mathrm{e}^{-\frac{t}{\tau_c}}\right)$& $t^2-\frac{t^3}{3\tau_c}$ & $2\tau_ct-2\tau_c^2$ \\ \hline
         Spin Echo & $\begin{cases}1 \qquad 0\leq t'<t/2\\
    -1 \quad t/2\leq t'<t\end{cases}$& $2\tau_ct-2\tau_c^2\left(3+\mathrm{e}^{-\frac{t}{\tau_c}}-4\mathrm{e}^{-\frac{t}{2\tau_c}}\right)$ & $\frac{t^3}{6\tau_c}$ & $2\tau_ct-6\tau_c^2$ \\\hline
XY-8 & $\begin{cases}1 \qquad (m-\frac{1}{4})\tau_p\leq t'<(m+\frac{1}{4})\tau_p\\
    -1 \quad (m+\frac{1}{4})\tau_p\leq t'<(m+\frac{3}{4})\tau_p\end{cases}$ & $\frac{\tau_p^2}{12\tau_c}t$ & $\frac{\tau_p^2}{12\tau_c}t$ & $\frac{\tau_p^2}{12\tau_c}t$ \\
    \hline
     \end{tabular}
    \caption{Expressions of $\eta(t';t)$ and $\chi(t)$ for Ramsey/DEER, spin echo, and XY-8. In XY-8, we assume the inter-pulse spacing $\tau_p\ll\tau_c$. }
    \label{tab:chi}
\end{table*}

\subsection{Average over positional randomness}

Until now, we have considered only the role of dynamical fluctuations in our analysis of the decoherence. 
In this subsection, we review the effect of positional disorder on the observed decoherence decay.
Following Ref.~\cite{fel1996configurational}, we compute the positional disorder average by first considering $N$ system spins occupying a volume $V$ in $D$ dimensions, and explicitly performing the volume integration of each spin as follows:
\begin{equation}
\begin{split}
C(t) &= \int\cdots\int \frac{d^D\vec{r}_1}{V}\frac{d^D\vec{r_2}}{V}\cdots\frac{d^D\vec{r_N}}{V}\prod_{i=1}^{N}\exp\left\{-\frac{1}{2}\left[ \frac{J_z|g_i| \chi(t)^{\frac{1}{2}}}{2r_{i}^\alpha}\right]^2 \right\}\\
&= \left[ \int \frac{d^D\vec{r}}{V}\exp\left\{-\frac{1}{2}\left[ \frac{J_z|g| \chi(t)^{\frac{1}{2}}}{2r^\alpha}\right]^2 \right\}\right]^N = \left[1-\frac{1}{N}\frac{N}{V}\int d^D\vec{r}~\left(1-\exp\left\{-\frac{1}{2}\left[ \frac{J_z|g| \chi(t)^{\frac{1}{2}}}{2r^\alpha}\right]^2 \right\}\right)\right]^N.
\end{split}
\end{equation}

This last equality gives us the limit definition of the exponential ($\lim_{N\to\infty} (1+x/N)^N = e^{x}$) and, in the thermodynamic limit ($N,V\rightarrow \infty$ with fixed density $n\equiv \frac{N}{V}$), is given by:
\begin{equation}
\begin{split}
C(t) &= \exp\left\{-n\int d^D\vec{r}~\left(1-\mathrm{e}^{-\frac{1}{2}\left[ \frac{J_z|g|\chi(t)^{\frac{1}{2}}}{2r^\alpha}\right]^2 }\right)\right\} = \exp\left\{-n\int_0^\infty dr~\int d\Omega\left(1-\mathrm{e}^{-\frac{1}{2}z^2}\right)r^{D-1} \right\}\\
 &= \exp\left\{-\frac{n}{\alpha}\left(\frac{J_z\chi(t)^{\frac{1}{2}}}{2}\right)^{\frac{D}{\alpha}} \left[ \int_0^\infty dz \left(1-\mathrm{e}^{-\frac{1}{2}z^2}\right)z^{-\frac{D}{\alpha}-1} \right] \left[ \int d\Omega~ |g|^{\frac{D}{\alpha}} \right]  \right\} \\
 &= \exp\left\{-\frac{nDA_D}{\alpha}\left[-\frac{\Gamma(-\frac{D}{2\alpha})}{2^{\frac{D}{2\alpha}+1}}\right] \left[\frac{\overline{g}J_z\chi(t)^{\frac{1}{2}}}{2}\right]^{\frac{D}{\alpha}}\right\},
\end{split}
\label{eq:Sx_pos_ave}
\end{equation}
where we make the substitution $z= \frac{J_z|g|\chi(t)^{\frac{1}{2}}}{2r^\alpha}$, $A_D = \frac{\pi^{\frac{D}{2}}}{\Gamma(\frac{D}{2}+1)}$ is the volume of a $D$-dimensional unit ball, and $\overline{g} = \left(\frac{\int |g|^{\frac{D}{\alpha}}d\Omega}{\int d\Omega}\right)^{\frac{\alpha}{D}}$ is the averaged angular dependence over a $D$-dimensional solid angle. 

We note that the integral converges only when $2\alpha > D$. This condition captures the physical intuition that, for very long-range interactions (small $\alpha$), the probe spin interacts strongly with an extensive number of spins.
More precisely, the number of spins at distance $r$ from the probe increases as $r^{D-1}dr$, while their contribution scales as $r^{-2\alpha}$. The variance of the phase $\phi$, at any fixed time, is then given by $\int_0^R dr~r^{D-2\alpha-1}$, which precisely diverges with increasing $R$ when $2\alpha > D$.
In this regime, which does not apply for our measurements, the standard deviation of $\phi(t)$ becomes unbounded in the thermodynamic limit, the Gaussian approximation $C = e^{-\langle \phi^2\rangle/2}$ no longer applies, and a more careful analysis is required.
%which diverges for $2\alpha \le D$.

Combining the results in Table~\ref{tab:chi} and Eqn.~\ref{eq:Sx_pos_ave}, we obtain the analytical form of the decoherence signal as measured in our system. 
In both the short-time and the long-time limits, $\chi(t)$ has a simple power-law dependence on time $t$. 
In each limit, the form of the decay profile is a simple stretched exponential, from which we can also obtain the decay timescale as a function of the defect density. 
For example, for the early-time Ramsey decay, we have: 
\begin{align}
C^{\mathrm{Ramsey}}(t \ll \tau_c) &= \exp\left\{-\frac{nDA_D}{\alpha}\left[-\frac{\Gamma(-\frac{D}{2\alpha})}{2^{\frac{D}{2\alpha}+1}}\right] \left[\frac{\overline{g}J_zt}{2}\right]^{\frac{D}{\alpha}}\right\}
=\exp\left\{-\left[\left(-\frac{\Gamma(-\frac{D}{2\alpha})}{2^{\frac{3D}{2\alpha}+1}}\frac{DA_D}{\alpha}\right)^{\frac{\alpha}{D}} n^{\frac{\alpha}{D}}\overline{g}J_zt\right]^{\frac{D}{\alpha}}\right\}\\
&=\exp\left\{-\left(\frac{t}{T_2^{\mathrm{Ramsey}}}\right)^{\frac{D}{\alpha}}\right\},\notag
\end{align}
where 
\begin{equation}
T_2^{\mathrm{Ramsey}}=\left[\left(-\frac{\Gamma(-\frac{D}{2\alpha})}{2^{\frac{3D}{2\alpha}+1}}\frac{DA_D}{\alpha}\right)^{\frac{\alpha}{D}} n^{\frac{\alpha}{D}}\overline{g}J_z\right]^{-1}\propto n^{-\frac{\alpha}{D}}.
\end{equation}
Similarly, we can obtain the analytical forms for other pulse sequences and in different time regimes, with all results summarized in Table~\ref{TableProfile}.

\begin{table}
  \centering
 \begin{tabular}{c||c|c||c|c} 
    \hline 
    & \multicolumn{2}{c||}{Early-time $t \ll \tau_c$} & \multicolumn{2}{c}{Late-time $t \gg \tau_c$} \\ \hline  
     & Stretch power & Decay timescale & Stretch power & Decay timescale \\ 
    \hline\hline
    
    \makecell{Ramsey (DEER)} & 
    \makecell[c]{$\dfrac{D}{\alpha}$} & \makecell[c]{$\left(K\overline{g}n^{\frac{\alpha}{D}}J_z\right)^{-1}$} 
    & $\dfrac{D}{2\alpha}$ & $\left(\frac{1}{2\tau_c}\right)\left(K\overline{g}n^{\frac{\alpha}{D}}J_z\right)^{-2}$ \\[5pt]
    \hline
    \makecell{Spin Echo} & $\dfrac{3D}{2\alpha}$ & $\left(6\tau_c\right)^{\frac{1}{3}}\left(K\overline{g}n^{\frac{\alpha}{D}}J_z\right)^{-\frac{2}{3}}$ & $\frac{D}{2\alpha}$ & $\left(\frac{1}{2\tau_c}\right)\left(K\overline{g}n^{\frac{\alpha}{D}}J_z\right)^{-2}$ \\[5pt]
    \hline
    XY8 & $\dfrac{D}{2\alpha}$ & $\left(\frac{12\tau_c}{\tau_p^2}\right)\left(K\overline{g}n^{\frac{\alpha}{D}}J_z\right)^{-2}$ 
    & $\dfrac{D}{2\alpha}$ & $\left(\frac{12\tau_c}{\tau_p^2}\right)\left(K\overline{g}n^{\frac{\alpha}{D}}J_z\right)^{-2}$ \\[5pt] 
    \hline
  \end{tabular}
  \caption{Ensemble averaged decay profiles for Ramsey/DEER, spin echo, and XY-8 pulse sequences; $\overline{g}$ is the averaged angular dependence, and $K=\frac{1}{2}\left[-\frac{DA_D}{\alpha}\frac{\Gamma(-\frac{D}{2\alpha})}{2^{\frac{D}{2\alpha}+1}}\right]^{\frac{\alpha}{D}}$ is a dimensionless constant only depending on $D$ and $\alpha$. 
    }
    \label{TableProfile}
\end{table}

Here, let us emphasize the role of positional disorder in determining the shape of the decoherence decay profile, as highlighted by the differences between Eqns.~\ref{eq:Sx2} and \ref{eq:Sx_pos_ave}.
% As highlighted by the difference between  (excluding positional averaging) and Eqn.~ (including positional averaging), the decay associated with a single spatial configuration is qualitatively different from the decay after averaging over positional disorder. 
%At its core, this is because,
To explain this difference, it is important to emphasize that the main contribution to the average coherence is dominated by different positional configurations at different times.
For example, at early times, only the configurations with very nearby spins are able to decohere, while configurations with distant spins have yet to contribute to the signal. 
Similarly, at late times, the configurations with fast decay rates have fully decohered, so the signal is dominated by configurations where the spins are far away from the probe and decoherence is slow.
This explains why the stretch power is determined by both dimensionality (which determines how many system spins can be close to the probe) and the exponent of the power-law interactions (which determines how fast those configurations decohere). 
% Put another way, there is no  single typical spatial configuration. 

By contrast, if the system spins are positioned on a regular lattice, the decay profile no longer exhibits a distribution of time scales and thus follows the shape of the single positional realization case [Eqn.~\ref{eq:Sx2}]. 
In particular, for a regular lattice, the spin echo always decays as a stretched exponential with a stretch exponent of $3$ (Table~\ref{tab:chi}), independent of both $D$ and $\alpha$.

\subsection{Dependence of coherence time on dimension}

Thus far, we have focused primarily on the functional form of the decoherence dynamics of the probe spin. Another important insight from Eqn.~\ref{eq:Sx_pos_ave}, also highlighted in Table~\ref{TableProfile}, is that the dimensionality $D$ also affects the scaling of the coherence time with density as $T_2 \sim n^{-\alpha/D}$, where $n \equiv N/V$.\\
\\
Naively, for fixed $\alpha$, this scaling may appear to imply that the coherence time in lower-dimensional samples decays faster than in higher-dimensional samples.
However, to directly compare the coherence time across dimensions, one must find a setting where the different densities (which carry different units) can be related.  
In practice, our delta-doped layer provides such a setting; namely, because it has a finite thickness $w$, we can relate the two-dimensional areal density to the three-dimensional volume density $n_{\mathrm{2D}}=\ n_{\mathrm{3D}}\cdot w$.
At low densities $n_{\mathrm{3D}}^{-1/3} \gg w$, the thin layer can be treated as a two-dimensional system.
Within this setting, the coherence time of the probe spin in the two-dimensional layer will be longer, since it is surrounded by a much smaller number of system spins.
Thin layers which do not satisfy this condition, i.e. which contain large spin densities $n_{\mathrm{3D}}^{-1/3} \leq w$, can no longer be considered two-dimensional and recover the three-dimensional scaling.
% However, as the density is increased, the spin-spin distance decreases as $n^{-1/D} \sim n_{\mathrm{3D}}^{1/D}$, and thus the coherence time decreases faster in the lower dimension setting.
% This does not imply that the coherence time in the two-dimensional layer eventually becomes smaller than three-dimensional setting; as the spin-spin spacing $n_{\mathrm{3D}}^{-1/3}$ becomes smaller the thickness of the two-dimensional layer, the layer can no longer be treated as a two dimensional system and the probe spin is surrounded by system spins all around---the layer now hosts a three dimensional spin ensemble.

% That is to say, instead of having a single $\tau_c$, one should consider a distribution of $\tau_c$, and obtain an average $\chi(t)$ according to such distribution. 
% Fortunately, at both early times ($t\ll \tau_c$) and late times ($t\gg \tau_c$), this ensemble average will not change the stretch exponents, and the $\tau_c$ in the expression should be understood as its ensemble-averaged value. 

\section{Quantum description}
\label{sec:quantum}
In Sec.~\ref{sec:Gauss-Markov}, we assumed that the spin dynamics can be captured by classical Gaussian random variables.
In general this need not be true, and one should treat the spin dynamics quantum-mechanically.
The goal of this section is to perform such an analysis and use it to highlight how the distinction between Gauss-Markov and telegraph noise can be cast as a property of the scrambling dynamics of the system spin operators $\hat{s}^z_i(t)$. 

% We now turn to a full quantum mechanical description which will allow us to understand the distinction between Gauss-Markov and telegraph noise.

The full Hamiltonian $H$ governing the dynamics in our experiments %of our setup 
is composed of the Ising interaction $H_z$ between the probe spin and the systems spins, as well as $H_s$ which determines the interactions between the system spins. Including the effect of the pulse sequence yields
\begin{equation}
    H_\text{tot}(t') = \eta(t';t) H_z +H_\mathrm{s}.
\end{equation}
%Second, the system no longer exhibits independent classical trajectories (unless those arising from our polychromatic drive protocol---we return to this point later). 
The initial density matrix of the system immediately encodes the ensemble of possible initial states.
In particular, we consider initial state of the system to be a fully mixed state for the system's spins, while the probe spin is initialized in the $\hat{x}$ direction. This corresponds to the following density matrix:
\begin{equation}
    \rho(t=0) = \frac{\mathds{1}}{\mathcal{N}_s} \otimes \left( \frac{\ket{\uparrow} + \ket{\downarrow}}{\sqrt{2}} \right) \left( \frac{\bra{\uparrow} + \bra{\downarrow}}{\sqrt{2}} \right),
\end{equation}
where $\mathcal{N}_s$ is the size of the Hilbert space of the spin system.

The probe coherence then becomes:
\begin{equation}
    C(t) = 2\mathrm{Tr} \left[\hat{s}^x_p U(t) \rho U^\dag (t)\right],\quad \mathrm{where}\quad U(t) = \mathcal{T}\exp\left[-i \int_0^t dt' H_{\mathrm{tot}}(t')\right].
\end{equation}

Because the dynamics conserve $\hat{s}^z_p$, the unitary operator $U$ can be divided into two operators $U_{\uparrow/\downarrow}$, which act only on the system spins and compute their evolution conditioned on the state of the probe:
\begin{equation}
    U(t) = U_\uparrow(t) \otimes \ket{\uparrow} \bra{\uparrow} + U_{\downarrow}(t) \otimes  \ket{\downarrow}\bra{\downarrow},
\end{equation}
% \begin{
%     U(t)\left[ \ket{\psi} \otimes \left(a_\uparrow \ket{\uparrow} + a_{\downarrow}\ket{\downarrow}\right) \right] = a_\uparrow \left[ U_\uparrow(t) \ket{\psi}\right] \otimes \ket{\uparrow} + a_\downarrow \left[ U_\downarrow(t) \ket{\psi} \right]\otimes \ket{\downarrow},
% \end{equation}
where
\begin{equation}
    U_\uparrow(t) = \mathcal{T}\mathrm{e}^{-i\int_0^t [H_\mathrm{s} + \eta(t';t)\sum_i\frac{J_zg_i}{2r_i^\alpha}\hat{s}^z_i]dt'},\qquad U_\downarrow(t) = \mathcal{T}\mathrm{e}^{-i\int_0^t [H_\mathrm{s} - \eta(t';t)\sum_i\frac{J_zg_i}{2r_i^\alpha}\hat{s}^z_i]dt'}.
\end{equation}

The coherence can then be written in terms of $U_{\uparrow / \downarrow}(t)$ acting on the spin system:
\begin{equation}
    C(t) = \frac{1}{\mathcal{N}_s}\mathrm{Tr}\left[U_\uparrow(t) U^\dag_\downarrow(t) + U_\downarrow(t)U^\dag_{\uparrow}\right] = \frac{2}{\mathcal{N}_s} \mathrm{Re}\left( \mathrm{Tr}\left[U_\uparrow(t) U^\dag_\downarrow(t)\right] \right)
\end{equation}

The dynamics $U_{\downarrow/\uparrow}(t)$ remain very complex because they include a contribution from both the system's dynamics and the probe-system coupling.
To understand how the probe spin is sensitive to the spin dynamics generated by the many-body interactions $H_s$, we move to the interaction frame of $H_s$ by making the following substitution:
\begin{equation}
    U_{\uparrow/\downarrow}(t) =U_0(t)\tilde{U}_{\uparrow,\downarrow} (t),\quad\mathrm{ with} \quad
    U_0(t) =  \mathrm{e}^{-i t H_\mathrm{s}},
\end{equation}
where
\begin{equation}
    \tilde{U}_{\uparrow/\downarrow} = \mathcal{T}\mathrm{e}^{\mp i\int_0^t\eta(t';t)\sum_i\frac{J_zg_i}{2r^\alpha_i}\hat{\tilde{s}}^z_i(t')dt'}, \quad \text{and}\quad
     \hat{\tilde{s}}^z_i(t) = U^\dag_0(t)\hat{s}^z_i(t)U^{}_0(t). 
     \label{eq:S19}
\end{equation}
In this frame, the two evolution operators lose an explicit reference to $H_s$ at the expense of the spin operators $\hat{s}^z_i$ becoming time-dependent. 
The resulting coherence remains in the same form, but with the unitaries now referring to the interaction frame: 
\begin{equation}
% \begin{split}
    % & \label{eq:single_contribution}\\
    C(t) = \frac{2}{\mathcal{N}_s} \mathrm{Re}\left( \mathrm{Tr}\left[\tilde{U}_\uparrow(t) \tilde{U}^\dag_\downarrow(t)\right] \right) 
% \end{split}
\label{eq:sigma_z_trace_int}
\end{equation}

% \begin{equation}
% % \begin{split}
%     % & \label{eq:single_contribution}\\
%     C(t) = \frac{2}{\mathcal{N}_s} \mathrm{Re}\left( \mathrm{Tr}\left[\tilde{U}_\uparrow(t) \tilde{U}^\dag_\downarrow(t)\right] \right) = \frac{2}{\mathcal{N}_s} \mathrm{Re}\left( \mathrm{Tr} ~ \tilde{U}_\uparrow^2(t) \right)
% % \end{split}
% \label{eq:sigma_z_trace_int}
% \end{equation}

Note that Eqn.~\ref{eq:sigma_z_trace_int} is formally similar to Eqn.~\ref{eq:Sx1}. 
Especially, replacing the quantum operator $\hat{\tilde{s}}^z_i(t)$ with a classical variable immediately reduce this equation to Eqn.~\ref{eq:Sx1}.
%Note that Eqn.~\ref{eq:sigma_z_trace_int} is formally the same as Eqn.~\ref{eq:Sx1}, except that $\hat{\tilde{s}}^z_i(t)$ is now a quantum operator rather than a classical variable.

% The full quantum expression for the probe spin coherence $\braket{\hat{s}_p^x}$ obtained in Eqn.~\ref{eq:sigma_z_trace_int} provides several important insights into the semi-classical approach we used in the main text and in the previous sections. 
% First, Eqn.~\ref{eq:sigma_z_trace_int} is formally the same as Eqn.~\ref{eq:Sx1}, except that $\hat{\tilde{s}}^z_i(t)$ is now a  quantum operator. Second, assuming that different spins $\hat{\tilde{s}}^z_i(t)$ are independent, the problem reduces to an evaluation of the eigenvalues of the single spin evolution operator $\tilde{U}_{\uparrow}$ for each spin $i$ independently.

% \fm{What are we trying to emphasize with these two insights?}
% \bty{Maybe the two points we want to emphasize are: 1) Eqn.S3 is simply replacing the quantum operator with a continuous classical variable. 2) By assuming the independence between different spins, the problem can be simplified. The second point may not be important. }

% given by
% \begin{equation}
%     \mathcal{U} = \mathcal{T}\mathrm{e}^{-i\int_0^t \eta(t')\frac{J_zg_i}{2r_i^\alpha}\hat{\tilde{s}}^z_i(t')dt'}.
% \label{eq:single_contribution}
% \end{equation}

While Eqn.~\ref{eq:sigma_z_trace_int} already averages over the possible initial states of $\hat{s}_i^z$, one must independently average the signal over the different spin ensembles (which arise, for example, from different positions of the spins, coupling to a polychromatic driving field or to other classical fluctuating degrees of freedom, etc.) % \sout{positional and on-site disorder,} etc.).
%\sout{This last point about two kinds of averages (one from different configurations of the many-body system and the other from the randomness of $H_\mathrm{s}$) is essential for determining whether a telegraph or a continuous (Gaussian) random variable can better describe the many-body noise of $\hat{s}^z_i$.
%We note that, since the auto-correlator $\xi(t) \propto \braket{ \hat{\tilde{s}}^z_i(t)\hat{\tilde{s}}^z_i(0)}$ always performs the two types of averages simultaneously, it does not contain the full information of the many-body noise. }
While these two kinds of averages (one from different configurations of the many-body system and the other from the randomness of $H_\mathrm{s}$) can lead to the same auto-correlator $\xi(t) \propto \braket{ \hat{\tilde{s}}^z_i(t)\hat{\tilde{s}}^z_i(0)}$, they are essential for determining higher order moments of the distribution and, thus, whether a telegraph or a continuous (Gaussian) random variable is a good description of the many-body noise of $\hat{s}^z_i$ and thus the decoherence signal of the probe spin. 
% \fm{Not sure if this sentence is the best way of emphasizing this point.}

% \fm{Can we simplify this discussion to just saying, look S18 is wonderful but you have to average positional disorder on top of it. Is there a deeper point we're trying to transmit?}
% \bty{I rethought about this part, and realized that "positional disorder" is misleading here. The two kinds of noise are bath trajectories and randomness in H. And in particular, we should mean temporal randomness in H.}

%one cannot distinguish between the two different noise models by just measuring $\xi(t)$,  %and will only get the same result. 

\subsection{Understanding decoherence dynamics in different physical scenarios}

While Eqn.~\ref{eq:sigma_z_trace_int} provides the generic expression for the decoherence dynamics of a spin coupled to a dynamical system, it is intractable to directly compute the decoherence profile from this equation except in specific cases. 
Nevertheless, such solvable points can provide important intuition for the conditions under which Eqn.~\ref{eq:sigma_z_trace_int} reduces to a semi-classical description with either Gauss-Markov or telegraph noise. 
In the following subsections, we provide two instructive examples where an explicit computation can be performed, and the relationship between the nature of the system  and its noise properties is made clear. 
%performing the necessary computation is intractable except in specific cases.
%In the following subsections, we describe two instructive examples where the explicit computation can be performed and the relationship between the nature of the bath and the Gaussian or Telegraph noise is made clear. 

\subsubsection{Probe coupled to a single spin evolving under an external drive}
First, we consider the case where the noise that decoheres the probe is generated by a single system spin, whose dynamics are controlled by an \emph{external random} drive.
% To better illustrate the two qualitatively different outcomes, we consider two simple but illustrative scenarios: (i) an isolated spin subject to the polychromatic drive described in the main text, which gives rise to continuous Gaussian noise; and (ii) a spin undergoing phonon-induced decay/spontaneous emission, which  gives rise to telegraph noise. The interacting many-body case lies somewhere between these two extremes. 
In this case, the interaction Hamiltonian is given by:
\begin{equation}
    H_\mathrm{s} = \Omega [ \hat{s}_i^x  \cos \theta(t)+ \hat{s}_i^y  \sin\theta(t)  ]
\end{equation}
where $\Omega$ characterizes the strength of the drive and $\theta(t)$ is a time-dependent phase.
The presence of such a time-dependent phase mirrors to the polychromatic drive described in the main text---$\theta(t)$ is chosen to follow a Gaussian stochastic process (see Methods, Sec.~IV.4) ~\cite{joos2021protecting}, and is randomized across different runs of the experiment. 
Crucially, for each run of the experiment, the dynamics induced by $H_\mathrm{s}$ generate a particular trajectory around the Bloch sphere without any loss of single-particle coherence.
As a result, the continuous spin rotation leads to a continuous change in the strength of the noise generated---this leads to the natural description of $s^z_i(t)$ as a continuous classical variable.

We emphasize that, within this framework, there is a single phase accumulated due to the noise for the particular driving $\theta(t)$. 
As a result, to obtain Gaussian-distributed noise, one must additionally average over different instances of $\theta(t)$.
In the experiment this corresponds exactly to the polychromatic drive, where the phase accumulation rate in each experimental run is random and changes continuously in time.\\

% is determined by a different $\Hint$ and the spin itself remains completely coherent on a single-particle level throughout the measurement duration $t$. We can picture the polychromatic drive as performing random rotations of the spin vector on the Bloch sphere; equivalently, from the perspective of operator evolution, the spin's $z$-component $\hat{\tilde{s}}^z_i$ continuously evolves in its two-dimensional Hilbert space. Intuitively, this leads to a continuous model for $s_i^z(t)$. 
% While each run of the spin dynamics only results in a value of single phase accumulation --- the evolution operator $\tilde{U}_{\uparrow,\downarrow}$ only has a pair of eigenvalues $\mathrm{e}^{\pm i\phi}$ --- the temporal randomness in $\Hint$ leads to a distribution of $\phi$ over multiple runs of the experiment. 

\subsubsection{Probe coupled to a system strongly interacting with a large bath}

We now turn to the opposite limit, where the noise is generated by spins coupled to a Markovian bath. 
The dynamics of $\hat{s}_i^z(t)$ can be thought to undergo spontaneous emission and absorption of photons/phonons---starting in either the state $\ket{\uparrow}$ or $\ket{\downarrow}$, the system undergoes quantum jumps into the opposite state at a rate given by $1/\tau_c$~\cite{meystre:2007}.
In this intuitive picture, the decoherence of the probe spin should be evaluated by averaging over all the possible quantum jump trajectories of the system spins---this precisely corresponds to a telegraph-like classical noise.

We can make this picture more precise, using Eqn.~\ref{eq:S19} as well as intuition from the perspective of operator dynamics. 
In particular, if the operator $\hat{\tilde{s}}_i^z(t)$ spreads across a large number of degrees of freedom much faster than the interaction time scale with the probe spin, then the probe spin is interacting with independent, commuting operators at different times, $[\hat{\tilde{s}}_z^i(t),\hat{\tilde{s}}_z^i(t')]=0$.
% As a result, the operators at different times commute with each other, i.e. $[\hat{\tilde{s}}_z^i(t),\hat{\tilde{s}}_z^i(t')]=0$. 
This immediately leads to two consequences: (i) the time-ordering operator in Eqn.~\ref{eq:S19} acts trivially on the exponential, and the eigenvalues of the exponential are simply given by the exponential of the eigenvalues of $\int_0^t \eta(t')\hat{\tilde{s}}^z_i(t')dt'$;
(ii) $\hat{\tilde{s}}^z_i(t)$ can be simultaneously diagonalized for different times. 

The latter fact enables a very simple analysis of the dynamics. 
To see this more easily, let us simplify our problem by dividing the time evolution into $M$ independent blocks of duration $\delta t = t/M$:
\begin{equation}
    \tilde{U}_{\uparrow/\downarrow} = \prod_j \exp \left [ \mp i \eta(j\delta t; t) \sum_i \frac{J_z g_i}{2r^\alpha_i} \hat{\tilde{s}}_i^z(j\delta t)\right],
\end{equation}
where $\hat{\tilde{s}}_i^z(j\delta t)$ commute with one another for different $i$ and time $j\delta t$.
Because there is a single eigenbasis $\ket{\mu}$ that diagonalizes all these operators, when computing $\bra{\mu}\tilde{U}_{\uparrow/\downarrow}\ket{\mu}$, the operators can be substituted by the corresponding eigenvalues $\lambda^i_{\mu}(j\delta t)$, and the exponent simply becomes the sum of individual contributions.
% In this common eigenbasis, each eigenvector, labelled by $\mu$, has a time-dependent eigenvalue $\lambda_{\mu}(t)$ corresponding to the time-dependent operator $\hat{\tilde{s}}^z_i(t)$, and eventually contributes an eigenvalue $\int_0^t\eta(t')\lambda_{\mu}(t')dt'$ to the spectrum of $\int_0^t \eta(t')\hat{\tilde{s}}^z_i(t')dt'$. 
Crucially, since $\hat{s}_z^2(t)=1/4$, $\lambda^i_{\mu}(t)$ can only be $\pm1/2$ \footnote{As long as the eigenspectrum of the original local operator is discrete, so will $\lambda_{\mu}(t)$; as a result, we expect the same telegraph noise description in the context of higher spin systems.}. 
The last necessary ingredient is the understanding of how $\lambda^i_{\mu}(j\delta t)$ at different times relate with one another.
In general, this depends on the details of the dynamics.
But, owing to the size of the Hilbert space, $2\mathcal{N}_s$, one can take a statistical approach to this question.
Namely, over the ensemble of eigenstates and operators, changing the sign of the eigenstate is expected to follow a Poisson process with time scale $\tau_c$,  $\langle \lambda^i_\mu(t)\lambda^i_\mu(t')\rangle\propto \mathrm{e}^{-|t-t'|/\tau_c}$.
We note that this correlation decay can be exactly recast as an exponential decay correlation function for the spin operators.
The dynamics of each $\lambda^i_\mu(t)$ can then be understood as either a single quantum jump trajectory (in the quantum language), or a single classical telegraph noise realization (in the classical description).
Summing over all the eigenstates $\ket{\mu}$ completes the trace operation in Eqn.~\ref{eq:sigma_z_trace_int} and yields the final coherence of the probe spin.

Here, we hasten to emphasize that unlike the single driven spin case, one does not have to compute an average over different trajectories of the spin dynamics, that is already incorporated within the trace operation and the sum over the different eigenstates.
This contrasts with the single spin example, where an explicit averaging over the driving fields was necessary to obtain the distribution of accumulated phases. 

% In quantum language this is quantum jumps,
% In classical variable this is telegraph

% In this picture of operator spreading, each single eigenvector [with an eigenvalue of $\lambda_{\mu}(t)$] maps to a trajectory in the quantum jump formalism and also map to a realization of Lorenz-Markov stochastic process in the naive classical model with telegraph noise. 

\subsubsection{Spin coupled to a generic many-body interacting system}
Taking the above two examples into consideration, whether a generic many-body system is described by the Gauss-Markov or the Telegraph random variable is determined by the speed of the operator spreading. 
If the spreading of the operator is slow, the dynamics of $\hat{\tilde{s}}_z^i(t)$ remain constrained to a few sites throughout the measurement duration and the system appears coherent (leading to continuous Gaussian noise).
If the spreading of the operator is fast,  $\hat{\tilde{s}}_z^i(t)$ quickly spreads across many spins and the rest of the system acts as an effective Markovian bath, leading to telegraph noise. 

In our disordered, strongly-interacting system, we conjecture that disorder leads to the slow spread of $\hat{\tilde{s}}^z_i$, and the decay of the auto-correlator $\langle \hat{\tilde{s}}^z_i(0)\hat{\tilde{s}}^z_i(t)\rangle$ mostly results from the different trajectories of local dynamics (originating from different $H_\mathrm{s}$ owing to different initial configurations of the bath spins). 
This is consistent with our experimental observation of the spin-echo decay stretch power $\beta=3D/2\alpha$ for a three-dimensional dipolar ensemble, and is characteristic of the Gaussian noise model.

\bibliography{references}